# Using Ecometric Data to Explore Sources of Cross-Site Impact Variance in Multi-Site Trials

David R. Judkins and Gabriel Durham

Principal Scientist and Statistician

Abt Associates

David_Judkins@AbtAssoc.com

301-347-5952

Acknowledgments: This research was funded by the Office of Planning, Research and Evaluation within the Administration for Children and Families as part of their Career Pathways Research Portfolio. https://acf.hhs.gov/opre/project/career-pathways-research-portfolio.

Revised draft 11/7/2021

Using Ecometric Data to Explore Sources of Cross-Site Impact Variance in Multi-Site Trials

Abstract. A new method is proposed to explore sources of cross-site impact variance in multi-site trials of social interventions. With this approach, aggregate reports from participants in the treatment arm about the treatment experience are used to define potential mediators and aggregate reports from participants in the control arm about the outcome of interest are used to remove confounding due to unmeasured local contextual factors. Particular attention is paid to correcting bias due to measurement error in the mediator. Asymptotic theory for the validity of this approach is provided. Two example applications are given, one simulated and one empirical.

## 1. Introduction

In 2003, Bloom, Hill and Riccio[1] demonstrated that it is possible, in the context of a large collection of compatible randomized trials of social interventions, to convincingly model some of the variation in impacts across sites in terms of variables that might be manipulable in future interventions. This article was highly praised when published (e.g., receipt of the APPAM Vernon award for best article in JPAM) and became a chapter in the popular book, *Learning More from Social Experiments,* by Howard Bloom (2005). More recently, Weiss, Bloom and Brock (2014) formalized a framework for research of this type and suggested terminology. Consistent with that framework, we refer to malleable features of local treatments (both design features and implementation features) that lead to variation in local impacts as "mediators." This framework also stresses the importance of local context in explaining variation in local impacts, but does not give these features a concise name. We refer to them as "contextual factors." The goal of this type of research is to identify the mediators and unconfound their influence from those of contextual factors.

In the first author's personal experience, federal evaluation research directors are very interested in applying BHR-style methods to evaluations of federal anti-poverty and education programs. Nonetheless, there have been few published applications.[2] The paucity of applications is due to the difficulty of assembling a large collection of experiments with the same goals (30 at a bare minimum). Even when a large collection of experiments with the same goal is created (most likely via a federal intervention or grants program) with consistently measured outcomes, it can be difficult and expensive to also measure mediators and contextual factors consistently. In addition to using information about any planned variation in the intervention, a common approach to mediator measurement is to collect information about the implementation from the people who are responsible for the implementation, either as front-line staff (street-level bureaucrats in the coinage of Lipsky, 1980) or as local directors for implementation.[3] The standard approach to measuring contextual factors is to use public data about the locality (cf. Johnson, et al., 2016, in the context of a collection of RDD studies), but these public data may not be specific enough to remove confounding.

[1] Hereafter referred to as BHR.
[2] See Weiss, et al. (2017) for a summary of relevant research on education and training programs. See Morris, et al. (2018) for a prominent example in early childhood education.

This paper suggests an "ecometric" approach to measuring both mediators and contextual factors, whereby participant-level data are aggregated to the site level.[4] More specifically, data from treated subjects about the treatment experience are aggregated to serve as mediators, and outcome data from control subjects are aggregated to serve as contextual factors.[5] The participant data can be collected either by follow-up survey or harvested from administrative databases. If a follow-up sample is required to collect outcome data, it might only be a small marginal cost to also collect data about treatment experiences to serve as the mediators. Relevant dimensions of the treatment experience might include service receipt, the quality of interactions with intervention staff,[6] and early progress. The justification for including site averages of early progress as mediators is that early progress should be more malleable than long-term outcomes.

---

[3] Surveys of local directors are less expensive than surveys of front-line staff, but BHR established that aggregated staff data were better predictors of site-specific impacts than those based on local directors.

[4] The term "ecometrics," not to be confused with the much broader field of econometric analysis, was coined by Raudenbush and Sampson (1999) and picked up by Hox (2010) to refer to the use of average participant responses in multi-level models.

[5] BHR also used follow-up survey data on study participants to define program-level variables, but their concept was different. They created program-level mean contrasts between the treatment and control arms as measures of local treatment contrast. In this paper, we are suggesting the use of program-level means of just the treatment arm as measures of fidelity.

[6] There is a strand of research starting from Anderson (1970) and continuing through Davis (2015) that average student perceptions can be used to measure classroom climate in ways that can help explain individual learning.

Section 2 reviews past work, both on analytic approaches for understanding cross-site impact variability in multi-site trials and on the use of ecometric data in the social sciences. Section 3 discusses how to combine these two strands of research to use ecometric data to model variation in local effects, assuming that there is only one mediator of program impacts.[7] Section 4 contains two example applications, one simulated and one empirical. The discussion in Section 5 closes the paper. An appendix includes proofs of asymptotic approximations, and an on-line supplement contains SAS code to implement the approach. We hope that this ecometric approach can lead to a larger number of successful efforts to follow Bloom's admonition to "learn more from social experiments." Of course, the main obstacle to this line of research remains assembling a large collection of compatible local experiments. We do not recommend applying the methods in this paper to collections of fewer than 30 local experiments.

## 2. Past Work

This section first reviews the traditional approach to BHR-style analyses, and then reviews the research on ecometric data. First, however, we note that although the work of Reardon and Raudenbush (2013) and Reardon et al. (2014) also involves the use of multi-site randomized trials to study questions of mediation, their work has almost nothing in common with the work reported here. That literature concerns the use of site-by-treatment interactions as multiple instruments when using an instrumental-variables approach to estimate the effect of a mediator or multiple mediators on an outcome. In their work, the mediator can be at the person or site level and exclusion assumptions are made. This paper considers only site-level mediators and does not make an exclusion assumption.

[7] Although we think that the approach in this paper could be generalized to handle multi-dimensional mediation, this paper only deals with the situation of a single mediator. If there are other mediators, none of the results in this paper remain valid.

## 2.1 Experimental Design and Statistical Models for BHR-Style Analyses

Assume that there is a collection of sites where a service is provided on an experimental basis to people who win a local lottery for a limited number of program slots. Members of the control arm are free to seek similar services elsewhere in the community, but they generally have minimal contact with the organization providing the experimental service after randomization. There is a global model for the intervention, but each local site will have a director who may have the freedom (and may have been encouraged) to make limited modifications that they think will improve the effectiveness of the intervention in their local area for the type of people who are likely to be attracted to their program. The local director also typically hires the staff to directly serve the public. Their skills, energy, and attitudes will inject "idiosyncratic" variation into the implementation. Finally, local conditions influence the types of clients who present for services and their outcomes if they are assigned to the control arm. The planned variation is certainly malleable, the idiosyncratic variation by front-line staff is indirectly malleable (through hiring, training, and management procedures), and the local conditions are immalleable. All three factors—planned variation in services, staff-driven variation in implementation, and local conditions—can cause variation in local effects. Covariation of the first two factors with local impacts is referred to as "mediation," while covariation of the third factor with local impacts is referred to as "moderation."

In past studies such as BHR and Walton, Harvill and Peck (2019), interviews with local implementation directors were conducted to measure planned variation in program design. Additionally, interviews with local staff were conducted and interview data regarding how they served treatment participants was aggregated to measure the idiosyncratic variation introduced by local staff. As previously mentioned, public data about the locality (such as the local unemployment rate from the Bureau of Labor Statistics) were used to measure the third set of factors. In addition, data about participants is collected, either through a follow-up survey or via administrative dataset such as tax records.

Then a multi-level model for the follow-up outcome given a single mediator is[8]

$$y_{ij} = \alpha + \left( X_{ij} - \bar{X}_{.j} \right)\beta + B_j\theta + L_j\xi + T_{ij}\left( \delta + B_j\varphi + L_j\eta + Z_j\lambda + q_j \right) + u_j + e_{ij} \qquad (1)$$

$$\begin{pmatrix} \beta_j \\ u_j \\ q_j \end{pmatrix} \sim \mathrm{N}\left( \begin{pmatrix} \beta \\ 0 \\ 0 \end{pmatrix}, \begin{bmatrix} [0] & [0] & [0] \\ [0] & \tau_u^2 & \tau_{uq} \\ [0] & \tau_{uq} & \tau_q^2 \end{bmatrix} \right),$$

$$\begin{pmatrix} B_j \\ L_j \\ Z_j \end{pmatrix} \sim \mathrm{N}\left( \begin{pmatrix} 0 \\ 0 \\ 0 \end{pmatrix}, \begin{bmatrix} \Sigma_B & 0 & \Sigma_{BZ} \\ 0 & \Sigma_L & \Sigma_{LZ} \\ \Sigma_{BZ} & \Sigma_{LZ} & \Sigma_Z \end{bmatrix} \right)$$

$$e_{ij} \sim \mathrm{N}\left(0, \sigma^2\right),$$

$$T_{ij} \sim \mathrm{B}(1, 0.5),$$

where

all pairs of variables with unnamed covariances are assumed to be independent of each other,

- $y_{ij}$ is the outcome of interest for the *i*-th participant in the *j*-th site,
- $X_{ij}$ is a row vector of baseline participant-level characteristics and $\bar{X}_{.j}$ is a row vector of the site-level averages of the baseline participant-level characteristics,

[8] This model could be equivalently written as two separate models: a level-one model for student-level outcomes given site features and a level-two model for the joint distribution of site features.

- $\beta$ is the vector of coefficients for the baseline participant-level characteristics,[9]
- $B_j$ is an *unmeasured* (latent) contextual moderator (or row vector of multiple unmeasured contextual factors),
- $L_j$ is a row vector of *measured* contextual factors (possibly including elements or all of $\bar{X}_{.j}$ ),
- $Z_j$ is the single mediator of interest,
- $\theta$ is a vector of coefficients indicating the influence of unmeasured contextual factors under control conditions,
- $\xi$ is a vector of coefficients indicating the influence of the measured contextual factors on participant outcomes under control conditions,
- $T_{ij}$ is a binary treatment indicator (the study participant either won or lost in the lottery),
- $\delta$ is the expected effect of treatment across sites,
- $\varphi$ is a vector of moderation effects caused by the unmeasured contextual factors,[10]
- $\eta$ is a vector of moderation effects caused by the measured contextual moderator,
- $\lambda$ is the primary parameter of interest, reflecting the mediation by $Z_j$ of the effect of treatment on the outcome,
- $q_j$ is an iid random site-level disturbance of the effect of treatment (idiosyncratic cross-site impact deviation),

---

[9] One could also assume that this coefficient varies across sites, but such a model would not be identified.

[10] The components of $B_j$ with non-zero coefficients in $\varphi$ need not be the same as those with non-zero coefficients in $\theta$ .

- $u_j$ is the iid random site-level offset reflecting differences between sites under control conditions (idiosyncratic variation in outcomes under control conditions),
- $e_{ij}$ is a participant-level error,
- $\tau_u^2$ is the between-site idiosyncratic variance in outcomes under control conditions,
- $\tau_q^2$ is the cross-site impact variance,
- $\tau_{uq}$ is the covariance between idiosyncratic variation in outcomes under control conditions and idiosyncratic cross-site impact deviation,[11]
- $\Sigma_B$ is the between-site covariance matrix for the unmeasured contextual factors,
- $\Sigma_L$ is the between-site covariance matrix for the measured contextual factors,
- $\Sigma_Z$ is the between-site covariance matrix for the mediators,
- $\Sigma_{BZ}$ is the between-site covariance between the latent confounder and the mediators, and
- $\Sigma_{LZ}$ is the between-site covariance between the measured moderators and the mediators, and
- $\sigma^2$ is the residual variance at the student level.

Under the assumption that either the unmeasured contextual factors do not moderate the effect of treatment ($\varphi = 0$) or the unmeasured contextual factors are independent of the mediators ($\Sigma_{BZ} = 0$),the critical parameter of interest, $\lambda$ , in model (1) can be easily estimated with any of several multi-level modeling software systems.[12] However, if neither of these conditions is true, then the estimated mediation ( $\hat{\lambda}$ ) will be inconsistent due to the confounding of the unmeasured contextual factors.

[11] It could happen, for example, that the program is more effective in sites with worse outcomes under control conditions.

Moreover, if the combined number of mediators and contextual factors is large relative to the number of sites in the study (as is pretty much uniformly true), then one must be concerned about small-sample biases in the estimates of model parameters. It is very easy to overfit model (1). So there is a strong tension between adding enough contextual factors to unconfound relationships between mediators and outcomes, but not so rich that there are no degrees of freedom left for mediators. Our approach, as explained in Section 3.1, offers the possibility of using just a single degree of freedom for contextual factors.

## 2.2 Ecometric Data in Social Science

A central idea of this paper, namely to use ecometric data to model effect heterogeneity, is not entirely new. There have been several prior treatments of the use of aggregate level-one information to define level-two variables in multi-level analysis. See for example, Asparouhov and Muthen (2007), Hox (2010, Section 10.3), Goldstein (2011, chapter 14), Marsh, et al. (2012), Aguinis and Culpepper (2015), Schunck (2016), Mackenbach et al. (2016), and Martin et al. (2017). Following Raudenbush and Sampson (1999), Hox refers to analyses using this sort of data as 'ecometrics.' Schunck warns against the bias caused by small level-one sample sizes, but methods to remove these biases have been developed by Asparouhov and Muthen as well as March, et al. Also, in climate study, there is similar work by Amman, Genton, and Li (2010) who advocate something they call Attenuation Corrected Ordinary Least Squares (ACOLS). None of this prior work, however, specifically concerns ecometrics in the context of multi-site experiments. In particular, although MPLUS supports methods to consistently estimate the influence of site-features measured from the participant perspective (Marsch, et al., 2012), this work does not consider studies that involve randomized treatment and control arms. So the methods in the next section are new.

[12] Stronger conditions would be needed to obtain consistent estimates of the coefficients of $L_j$, and of course, it is impossible to estimate the coefficients of $B_j$ since these variables are unmeasured.

## 3. Analysis with Ecometric Measurements of Moderators and Mediators

Simply substituting ecometric measures of malleable and immalleable features into equation (1) leads to biased estimates of $\lambda$ . Section 3.1 indicates how to "control" on ecometrically measured contextual factors. Section 3.2 builds upon the results of Section 3.1 to estimate the mediating role of a single ecometrically measured mediator while controlling for econometrically measured contextual factors. Another way of understanding the organization is that 3.1 focuses on corrections for confounding of mediation by contextual factors while 3.2 focuses on corrections for attenuation in estimated mediation due to measurement error of the mediators.

This development is based on forming the design-based estimate of the local impact and then trying to model it in terms of available data. Based on equation (1), this estimated local impact can be written

$$\hat{\mu}_j = \delta + B_j \varphi + L_j \eta + Z_j \lambda + q_j + g_j \qquad (2)$$

where $\hat{\mu}_j$ is the design-based estimate of the local impact (the simple local difference between the treatment and control arms) and $g_j$ is a level-two model error. As we demonstrate below, this form is useful for carrying out the ecometric approach to measurement mediators and contextual factors.

## 3.1 Fitting Models with Control arm Outcomes as Moderators

BHR demonstrated that the impact of welfare-to-work programs was inversely related to the local unemployment rates. When the local unemployment rate is high, the impact of the local program is smaller. More recently, Rothstein and von Wachter (2017) caution that the generalizability of local experiments may be limited by variation in local economic conditions, giving the example that "… a reemployment bonus program may have a very different effect in a full-employment local economy than when the local area is in a recession … ."

Suppose, however, that there is an important unmeasured contextual moderator that cannot be explained from those data. (Important in the sense that $B_j\varphi$ and $\text{cov}\left(B_j, Z_j \mid L_j\right)$ are both large.) For example, given the BHR results for welfare-to-work programs, it seems reasonable to theorize that a sectoral training program is likely to be less effective in areas where unemployment among workers trained for that sector is high. Federal sources may not provide data that are sufficiently detailed with respect to both geography and occupation. Moreover, they can only measure unemployment by occupation among people who have worked in that occupation. They cannot include in the unemployment rate denominator potential workers with relevant training. In this case, the omission of the unmeasured contextual moderator could lead to biased estimates of $\lambda$ , the influence of the mediator on the impact. We realized that one way to at least partially remove this bias is to use the control arm mean outcome as an indirect measurement of the unmeasured contextual moderator. Under some conditions, it may be sufficient, in fact, to use the control arm mean outcome as the sole contextual moderator, thereby preserving degrees of freedom for multiple mediators.

More specifically, our suggestion is to enter $\bar{y}_{\bullet jC}$ (the average local outcome for the control arm) in place of the unobserved $B_j$ equation (2), resulting in:[13]

$$\hat{\mu}_j = \delta + \bar{y}_{\bullet jC}\varphi + L_j\eta + Z_j\lambda + q_j + g_j \tag{3}$$

Assuming that local sample size is large enough to ignore the site average of person-level residuals, we see from equation (1) that the average outcome in the local control arm $\bar{y}_{\bullet jC}$ can be approximated as

$$\begin{aligned} \bar{y}_{\bullet jC} &= \frac{1}{n_j}\sum_i \left(1 - T_{ij}\right) y_{ij} = \\ &= \frac{1}{n_j}\sum_i \left(\alpha + \left(X_{ij} - \bar{X}_{.j}\right)\beta_j + B_j\theta + L_j\xi + u_j + e_{ij}\right) \\ &\approx \alpha + B_j\theta + L_j\xi + u_j \end{aligned}$$

So, if $\text{corr}\left(B_j\theta + L_j\xi + u_j, B_j\varphi + L_j\eta\right)$ is large, then controlling on $\bar{y}_{\bullet jC}$ should remove much of the confounding of both measured and unmeasured contextual factors from the relationship of outcomes to mediators. In this case, one could actually drop $L_j$, thereby preserving degrees of freedom. Alternatively, if only $\text{corr}\left(B_j\theta + u_j, B_j\varphi\right)$ is large, then one may condition on both $\bar{y}_{\bullet jC}$ and *L*. In this case, the new information gained is $B_j\theta + u_j$. Adding the unmeasured $B_j$ would remove all confounding and would thus be the ideal but unattainable model. Using $\bar{y}_{\bullet jC}$ may be an acceptable substitute (in terms of removing confounding) if $B_j\theta$ is a sufficient statistic for $B_j$ (automatic if there is a single latent variable that influences both control arm outcomes and local impacts, but otherwise not guaranteed) and if local idiosyncratic variation in outcomes under control conditions is small compared to the influence of the unmeasured contextual moderator. Technically, we need

$$\text{var}\left(B_j\theta\right) = \theta'\Sigma_B\theta >> \tau_u^2 = \text{var}(u_j)\,.$$

[13] One can also consider dropping $L_j$ from this model, as discussed below.

If this is true, then controlling on $\bar{y}_{\bullet jC}$ should remove confounding almost as well as if we could somehow control on unmeasured local moderator directly. Another way to think about the condition that $\theta'\Sigma_B\theta >> \tau_u^2$ is to say that unmeasured systemic sources of variation in the outcome under control conditions are much more important than idiosyncratic sources. For example, if the outcome is welfare participation, one would require that unmeasured contextual factors like local public disapproval of welfare recipients are more important than idiosyncratic sources such as extreme weather events or locust infestations. If this is not true, one can still hope that $\varphi'\Sigma_B\varphi$ is also small relative to $\lambda'\Sigma_Z\lambda$, meaning that the bias in the estimated mediation effect is small. This is perhaps not unreasonable to assume. The logic is that if an unmeasured contextual feature has little influence on outcomes under control conditions, then it is implausible to assume that it will have a large influence on the impact of treatment.

Experimentation with alternative model fitting procedures led to the following three-phase process. In the first phase, regression-adjusted estimates of local impacts are obtained using only level-one covariates. In the second phase, best linear unbiased predictions (BLUPs) of local mean outcomes under control conditions are obtained.[14] In the third phase, the regression-adjusted estimates of local impacts modeled in terms of BLUPs of local mean outcomes, any measured contextual moderator(s), and the mediator(s) of interest to obtain estimates of $\lambda$.

*First phase*

In the first phase, fit the model:[15]

$$y_{ij} = \alpha_j + \left(X_{ij} - \bar{X}_{.j}\right)\beta + T_{ij}\delta + T_{ij}\gamma_j + e_{ij} \tag{4}$$

From this model, we obtain estimates of the local effects as

$$\hat{\mu}_j = \hat{\delta} + \hat{\gamma}_j \tag{5}$$

and obtain estimated conditional variances $S_j^2$ of $\operatorname{var}\left(\hat{\mu}_j \mid q_j, u_j\right)$ on these estimates. These can be noisy with small local sample sizes. To stabilize them, average then with the true local variance under the assumption of heteroscedasticity of level-one errors, $4\sigma^2/n_j$, where $n_j$ is the local sample size, across arms, and it assumed that the sample is split evenly between treatment and control conditions.[16]

---

[14] For an introduction to BLUPs, see Robinson (1991).
[15] If the number of participant-level baseline covariates is large, it might be wise to add a variable selection phase first, such as a LASSO.
[1616] If the probability of assignment to treatment has some value *p*, different than 0.5, then one would instead average with estimated conditional variance with $\sigma^2 / n_j / (p(1-p))$.

*Second phase*

In the second phase, fit the model:

$$\left(1-T_{ij}\right)y_{ij}=\alpha+\left(X_{ij}-\bar{X}_{.j}\right)\beta+L_j\xi+u_j+e_{ij} \tag{6}$$

From this model, obtain estimates of the BLUPs, $\hat{u}_j$, for the site-level random effects under control conditions. These are an unknown mixture of purely random effects and fixed effects due to the unmeasured latent confounder. Inclusion of $L_j$ is optional but should help improve the correlation of $\hat{u}_j$ with $B_j$. Obviously though one must be judicious in including level-two covariates in this phase since there are few available degrees of freedom.

*Third phase*

Finally, model the estimated local effects in terms of the BLUPs for local outcomes under control conditions, and the mediator in a single level model with two random error components:

$$\hat{\mu}_j=\delta+\hat{u}_j\varphi+L_j\eta+Z_j\lambda+q_j+g_j \tag{7}$$

where $\text{var}(q_j)=\tau_q^2$, $\text{var}(g_j)=(S_j^2+4/n_j)/2$ and *q* and *g* are iid errrors, independent of each other. The *q*-error reflects unexplained cross-site impact variability. The *g*-error reflects stabilized measurement error in the estimated local impact.[17] This model can be fit in

[17] As discussed earlier, we introduced the average of $4/n_j$ with $S_j^2$ to stabilize the variance estimates of the estimated local effects. The former is an approximate expected variance under an assumption of homoscedasticity. With small site sample sizes, GLIMMIX would frequently fail to converge if fed pure $S_j^2$.

SAS/GLIMMIX.[18] As discussed previously, we only expect this approach to be successful in removing confound by the unmeasured $B_j$ if $\theta'\Sigma_B\theta >> \tau_u^2$. Even in that circumstance the estimates of $\varphi$ and $\eta$ will not be interpretable. (The estimate of $\varphi$ will be bad because $\hat{u}_j$ is not an unbiased estimator of $B_j$. The estimate of $\varphi$ will also be bad because $\hat{u}_j$ is contaminated by $L_j$.)

## 3.2 Using Ecometric Measurements of Mediators

With regard to measuring local practices, this is closely related to the field of measurement of implementation fidelity. We cover two issues. First, what are the important dimensions of fidelity and second, how should they measured? The first issue is covered well by Weiss, Bloom, and Brock (2014). Both quantities and quality of services could matter, and services can be broadly defined. In the context of adult training programs (the impetus for our work), we think that the amount of training received could be an important mediator, but we focus in this paper on statistical methods, leaving the work of mapping the theory of program implementation to others.

Regarding measurement of fidelity, there is a modest literature on this this field, which has been summarized by James Bell and Associates (2009). There are three basic methods for fidelity measurement: self-report from intervenors, self-report from participants, and third-party observation (either in real time or via video recordings). The typical difficulty with third-party observation is that one cannot observe the intervenors for more than a small fraction of the time they spend with participants. Additionally, costs will be a driving issue as will be standardization of observers and the development of protocols that minimize disruption of the intervention by the observation process. Self-report from intervenors is a far less expensive approach but can still be expensive if there is a large pool of intervention staff, as might, for example, be the case if the intervention involves giving out training vouchers that can be used any of a large collection of training providers. Moreover, there can be issues with intervention workers giving socially acceptable responses, as well as limited self-awareness (Malone, 2015).

[18] This is not documented in SAS documentation. Our thanks to Pushpal Mukhopadhyay for showing us how to specify a complex variance structure for the "R-side" residuals in SAS/GLIMMIX. Using this procedure results in much smaller variances on the estimates of $\lambda$ than are obtained with OLS. Details are in supplemental online materials.

Self-report by participants is perhaps used the least frequently for fidelity measurement. Participant data on program quality is probably not useful.[19] However, in favor of using participants as informants, consider that participants continuously observe the intervenors during service sessions and might be very sensitive to the subconscious signals sent by intervention staff. Moreover, there is less need to be concerned about interrater reliability than with direct observation. By averaging the ratings of many students for a given intervenor, reliability is improved. Of course, if the number of survey participants per locality is small, measurement error again becomes an issue that will need attention in the analysis. Finally, if interviews are being conducted with participants for other purposes (i.e., outcome measurement), the marginal cost of the data will be low. The biggest challenge will be recall errors. To minimize this, two rounds of participant interviews may be required – one during service to report on local practices and one after service to collect outcomes. This would, of course, reduced the savings unless these interviews were already planned for other purposes.

[19] Smith, Whalley and Wilcox (2020) report no correlation between perceived impact and econometrically estimated impacts for participants in JTPA. Moreover, both Braga, Paccagnella and Pellizzari, 2014, and Carrell and West, 2010 report negative correlations between student evaluations of teachers and their subsequent readiness for learning in subsequent courses.

Assume that participant-level data are used to estimate the mediator with the following measurement error model:

$$z_{ij} = Z_j + \omega_{ij} \tag{8}$$

where $\omega_{ij} \sim \mathrm{N}\left(0, \sigma_{1Z}^2\right)$ are iid and independent of all other errors in equation (1).[20] Let $n_{jT}$ be the local treatment sample size and $\overline{z}_{.jT}$ be the average treated student report of the mediator within the site. Then the appropriate model for the third phase is:

$$\hat{\mu}_j = \delta + \hat{u}_j\varphi + L_j\eta + \overline{z}_{.jT}\lambda + q_j + g_j \tag{9}$$

where $\mathrm{var}\left(\overline{z}_{.jT}\right) = \sigma_{1Z}^2 / n_{jT}$, $\mathrm{var}(q_j) = \tau_q^2$, $\mathrm{var}(g_j) = (S_j^2 + 4/n_j)/2$ and $q$ and $g$ are iid errrors, independent of each other. Assuming temporarily that $\mathrm{var}\left(\overline{z}_{.jT}\right) = 0$, this model can also be fit in SAS/GLIMMIX. This however, results in an attenuated estimate of $\lambda$ , meaning that the expected value of the estimate is somewhere between 0 and its true value. We studied two ways to reverse this attenuation. One approach involves wrapping a SIMEX algorithm (Cook and Stefanski, 1994; Stefanski and Cook, 1995) around a SAS/GLIMMIX core. The other approach involves calculating a first-order approximation of the multiplicative bias and then dividing the biased estimate by the inverse of this estimated multiplicative bias. There follows a subsection devoted to each approach to removal of the bias caused by measurement error in the mediator.

[20] One might be able to relax the assumption of homoscedasticity in student-level errors across sites, but this is unclear.

**Using SIMEX to Remove Attenuation Bias**

The basic idea is to add more measurement noise to the estimated mediator and then extrapolate back to zero measurement noise on the mediator. Adding the measurement noise is the SIMulation step of SIMEX while extrapolating back to zero measurement noise is EXtrapolation step of SIMEX. No off-the-shelf software for SIMEX will handle the complexities of this adjustment in this context. However, it is not difficult to write a SAS macro to wrap around SAS/GLIMMIX. There are a variety of ways to execute each step. For the simulation step, we added extra noise $\varepsilon_{jb} \sim N\left(0, b(0.004)\hat{\sigma}^2_{1Z} / n_{jT}\right)$ to $\overline{z}_{.jT}$ , for *b*=1,…,500, so that at the extreme, measurement error on the mediator was doubled. Let $\hat{\lambda}_b$ be the estimate of $\lambda$ on the perturbed data using SAS/GLIMMIX, as discussed above. One expects that as *b* increases, $\hat{\lambda}_b$ should attenuate toward zero. For the extrapolation step, we fit the model

$$\hat{\lambda}_b = \pi_0 + \pi_1 b(0.004) + \pi_2 b^2 (0.004)^2 + e_b . \qquad (10)$$

The SIMEX estimate of $\lambda$ is

$$\hat{\lambda}_{SIMEX} = \hat{\pi}_0 - \hat{\pi}_1 + \hat{\pi}_2 . \qquad (11)$$

(This is motivated by at *b*=-250, the measurement error on $\overline{z}_{.jT}$ would be zero.)

The only known way to estimate the variance on $\hat{\lambda}_{SIMEX}$ is with a resampling procedure like the jackknife (Tukey, 1958; Efron and Stein, 1981; Wolter, 2007). This is computationally costly with 500 GLIMMIX runs per jackknife iteration. Decreasing the step size below 0.004 and increasing the upper bound of b so that measurement is still doubled at the extreme might improve SIMEX performance, but with the need to jackknife around that, 500 steps of size 0.004 seems reasonable.[21]

**Using Approximation of Expected Bias to Remove Attenuation Bias**

Under model (9) with the further simplifying assumptions that there is no moderation by contextual factors (i.e., $\varphi = \eta = 0$ )[22] and that the sample is evenly split between treatment and control in each site so that $\text{var}(w_j) \approx 4\sigma^2 / n_j$, we were able to derive several approximations of the attenuation in estimated mediation caused by using the ecometric approach to measuring the mediator.[23] All of these approximations assume that the sample sizes follow a gamma distribution. Beyond that, they vary in their assumptions about the relative magnitudes of the relative variance in the site sample sizes, $V_n^2$ and the idiosyncratic cross-site impact variance, $\tau_q^2$.

If *either* is negligible, then a reasonable disattenuation coefficient to apply to the estimate of $\lambda$ from model (9) is

$$\zeta_{(1)} = 1 + \frac{2(1-\hat{\rho}_Z)}{\bar{n}\hat{\rho}_Z}, \qquad (12)$$

where $\hat{\rho}_Z$ is the estimated intraclass correlation in Z and $\bar{n}$ is the average site sample size.

[21] The originators of SIMEX tested 1000 repetitions at each of 11 steps of width 0.10 (Stefanski and Cook, 1995). This many repetitions does not seem to be feasible when wrapping around GLIMMIX.
[22] This is not a realistic assumption, but it greatly simplifies the derivation and the simulation in Section 4.1 shows that adding moderation by both measured and unmeasured contextual factors did not affect the accuracy of these approximations.
[23] All derivations are in the appendix.

If $\tau_q^2$ is large but $V_n^2 < 1$, then the following slightly more complex disattenuation coefficient should work well:

$$\zeta_{(2)} = 1 + \frac{2\left(1-\hat{\rho}_z\right)\left(1+V_n^2\right)}{\overline{n}\,\hat{\rho}_z} \tag{13}$$

If $\tau_q^2$ is large and $V_n^2 \geq 1$, but still not very large, the following approximation might work better:

$$\zeta_{(2a)} = 1 + \frac{2\left(1-\hat{\rho}_z\right)\left(1+V_n^2-2V_n^4\right)}{\overline{n}\,\hat{\rho}_z} \tag{14}$$

Finally, if $\tau_q^2$ and $V_n^2$ are both large, then the following more complex disattenuation coefficient should be most accurate:

$$\zeta_{(3)} = 1 + \frac{2\left(1-\hat{\rho}_z\right)}{\overline{n}\,\hat{\rho}_z} \frac{\left(\tau_q^2 + 4\sigma^2/\overline{n}\right)^2 + \tau_q^4 V_n^2 - 2\tau_q^6 V_n^4 \left(\tau_q^2 + 4\sigma^2/\overline{n}\right)^{-2}}{\left(\tau_q^2 + 4\sigma^2/\overline{n}\right)^2 - 4\tau_q^2 V_n^2/\overline{n} + 8\tau_q^4\sigma^2 V_n^4 \left(\tau_q^2 + 4\sigma^2/\overline{n}\right)^{-2}} \tag{15}$$

One caution with this approximation is that special measures may need to be taken to bound the estimates of $\hat{\rho}_Z$ away from zero unless the number of sites is large. Some estimation approaches allow negative estimates and most allow zero estimates. Li and Lahiri (2010) has a method that guarantees positive estimates. It might also be possible to use external information about $\hat{\rho}_Z$ to calculate values of it for a variety of related potential mediators and then to average the individual estimates of intraclass correlation across the set of potential mediators.[24]

[24] A somewhat more sophisticated approach would be to model both total and within-site variances using the general variance functions discussed in Wolter (2007) and then use the modeled variance structures to estimate a collective intraclass correlation as in [citation deleted for blind review].

All four of these disattenuation coefficients tend to infinity as $\bar{n}\hat{\rho}_Z$ tends to zero. Depending on the size of the geographic area (such as whole metropolitan area or just a single classroom) represented by the site and the nature of the experience, intraclass correlation might easily range from 0.005 to 0.250,[25] so large local sample sizes will generally be required to avoid very large disattenuation coefficients.

Table 1 shows the values of the four disattenuation factors for different values of ICC in the mediator and local sample size, with all four making the same assumptions about relative variance in local sample sizes, the idiopathic component of cross-site impact variance, and the number of local sites. With an ICC on the mediator of 0.02 and an average site sample size of 20, the corrections are very large, but mild for an ICC on the mediator of 0.08 and an average site sample size of 150. The four factors are similar to each other across the range of scenarios, but $\zeta_{(1)}$ is generally close to $\zeta_{(3)}$ than the other two. Since $\zeta_{(3)}$ is based on the weakest assumptions and since $\zeta_{(1)}$ is the easier to calculate, this suggests that $\zeta_{(1)}$ may be the best choice.

[25] See Hedges and Hedberg 2007) for a compilation of ICCs in school contexts.

Table 1. Approximate Disattenuation Factors

| ICC in Mediator $\rho_Z$ | Average Local Sample Size $\bar{n}$ | Disattenuation Factor | | | |
|---|---|---|---|---|---|
| | | $\zeta_{(1)}$ | $\zeta_{(2)}$ | $\zeta_{(2a)}$ | $\zeta_{(3)}$ |
| 0.02 | 20 | 5.90 | 8.84 | 5.31 | 6.15 |
| 0.05 | 20 | 2.90 | 4.04 | 2.67 | 3.00 |
| 0.08 | 20 | 2.15 | 2.84 | 2.01 | 2.21 |
| 0.02 | 80 | 2.23 | 2.96 | 2.08 | 2.38 |
| 0.05 | 80 | 1.48 | 1.76 | 1.42 | 1.53 |
| 0.08 | 80 | 1.29 | 1.46 | 1.25 | 1.32 |
| 0.02 | 150 | 1.65 | 2.05 | 1.57 | 1.74 |
| 0.05 | 150 | 1.25 | 1.41 | 1.22 | 1.29 |
| 0.08 | 150 | 1.15 | 1.25 | 1.13 | 1.17 |

Notes: Based on $V_n^2 = .6$, $\tau_q^2 = 0.02$, $\sigma^2 = 1$, and $m = 40$

For variance estimation, some form of resampling like the jackknife seems the most practical since the need to bound $\hat{\rho}_Z$ away from zero seems intractable for a linearization approach to variance estimation. One advantage over these disattenuation factors over SIMEX is that the jackknife replications are far less computationally demanding, so it is feasible to run a larger number of jackknife replications.

## 4. Examples

This section includes two examples, one that is a "toy" simulation where truth is known but assumptions are not realistic and one that uses real data but where truth is not known. (Simplified models meant to illustrate basic principles are frequently referred to as toy models by mathematicians and statisticians.) The second is an application of the methodology to the same data previously analyzed by Peck et al (2018, chapter 7) and Walton, Harvill and Peck (2019).

### 4.1 Toy Simulation

This simulation is deliberately not realistic. Instead, it is designed as a demonstration that procedures can work under favorable circumstances. This is most apparent in the choice of the number of sites. A realistic number would be more in the range of 30-70, but this simulation assumes 500. We used this number because accurate simulation of the behavior of the methods on a collection of this size would have required an extra layer of simulation—something like 2000 draws, each with 30-70 sites instead of one draw with 500 sites. Such a simulation would be desirable since it could also evaluate the quality of the jackknifed variances, but the jackknifing of the SIMEX method was already very time intensive with a single draw. Imposing another layer of simulation would have required a different approach to computing.

In the toy example, there is a single unmeasured contextual moderator (*B*), a single measured contextual moderator (*L*) and a single mediator (*Z*) with the following distribution:

$$\begin{bmatrix} B_j \\ L_j \\ Z_j \end{bmatrix} \sim \mathrm{N}\left( \begin{bmatrix} 0 \\ 0 \\ 0 \end{bmatrix}, \begin{bmatrix} 1 & 0 & -.5 \\ 0 & 1 & 0 \\ -.5 & 0 & 1 \end{bmatrix} \right)$$

Note that the strong negative correlation between the unmeasured contextual moderator and the mediator sets up the possibility of strong confounding. In addition, there is a single baseline covariate at the person-level that is site-centered:

$$X_{ij} \sim \mathrm{N}(0, 0.1),$$

Additional population parameters are as follow:

$$\begin{bmatrix} \alpha \\ \beta \\ \theta \\ \xi \\ \delta \\ \varphi \\ \eta \\ \lambda \\ \rho_Z \\ \sigma^2 \end{bmatrix} = \begin{bmatrix} 1000 \\ 3.0 \\ 2.0 \\ -.5 \\ 0.8 \\ 0.5 \text{ or } 0 \\ 0 \\ 1.0 \\ 0.9 \\ 1.0 \end{bmatrix}$$

$$\begin{bmatrix} \tau_u^2 & \tau_{uq} \\ \tau_{uq} & \tau_q^2 \end{bmatrix} = \begin{bmatrix} 0.100 & -0.015 \\ -0.015 & 0.020 \end{bmatrix}$$

$$n_j \sim \Gamma(1,20)$$

$m$=500.

Putting some of these parameters into words, the overall effect of treatment is $\delta = 0.8$ standard deviations in the outcome, a rather large effect, a one-standard deviation increase in the unmeasured confounder $B$ intensifies the effect of the intervention by $\varphi = 0.5$ or 0 standard deviations in the outcome, a one-standard deviation increase in the roughly measured mediator Z intensifies the effect of the intervention by $\lambda = 1.0$ standard deviations in the outcome, the unmeasured contextual factor is strongly correlated with the mediator, the idiosyncratic variance in the outcome under control conditions is 10 percent of the total variance under control conditions, the idiosyncratic cross-site impact variance is small, just 0.02, the average student sample size per site is 20 with a relative variance of 1.0, and the total number of sites in 500.

The results are shown in Table 2. It contains two sets of results. The top panel shows results when then the unmeasured contextual factor does not moderate local impacts. The first row shows what would happen if the unmeasured contextual factor was somehow available and if the mediator was measured without error. This row is not achievable in practice but provides an upper bound on the quality of mediation estimation given the drawn sample. When there is no confounding and the mediator is measured without error, the estimated mediation of the effect of T caused by *Z* is 0.980 within sampling error of the true value. A naïve model that is fit without moderators and treating the mediator as if it were measured without error actually performs slightly better than this theoretical maximum (0.984 versus 0.980). Presumably this is because intraclass correlation in the mediator is high in this example and there is no confounding. None of the suggested corrections perform better than the naïve estimator, but their performance is generally not much worse than the naïve estimator.

The situation is very different in the lower panel where use of the naïve model yields a highly biased estimate of the mediating effect of *Z* (0.685 versus the truth of 1.000). Including $\hat{u}_j$ (the BLUP of local outcomes under control conditions) in the model led to a dramatic improvement in the estimate of the mediating effect of *Z* (0.922) with no increase in standard error relative to the naïve model. Adjusting for attenuation with SIMEX further improves the estimated mediation (0.938) but only modestly and with a sharp penalty in terms of computing time and standard errors. The various approximate disattenuation coefficients supply similar point estimates as SIMEX but with much smaller standard errors and much less computing time.

Table 2. Illustration of Methods on Large Toy Sample with and without Confounding

| Methodology | CBLUP in Model? | Method for Correction of Attenuation | $\hat{\lambda}$ (Mediating effect of Z) (Truth: $\lambda = 1$) | SE | Jack Reps | CPU Minutes |
|---|---|---|---|---|---|---|
| No confounding $\varphi = 0$ | | | | | | |
| *B* available, *Z* measured without error | | | 0.980 | 0.023 | | |
| Naïve model | No | None | 0.984 | 0.021 | | |
| GLIMMIX | Yes | None | 0.938 | 0.026 | | |
| GLIMMIX+ SIMEX | Yes | SIMEX (500) | 0.955 | 0.085 | 16 | 347 |
| GLIMMIX * Adj1 | Yes | Eq 12 | 0.954 | 0.029 | 64 | 4 |
| GLIMMIX * Adj2 | Yes | Eq 13 | 0.968 | 0.029 | 64 | 4 |
| GLIMMIX * Adj2a | Yes | Eq 14 | 0.943 | 0.029 | 64 | 4 |
| GLIMMIX * Adj3 | Yes | Eq 15 | 0.950 | 0.029 | 64 | 4 |
| Strong confounding $\varphi = 0.5$ | | | | | | |
| *B* available, *Z* measured without error | | | 0.980 | 0.023 | | |
| Naïve model | No | None | 0.685 | 0.029 | | |
| GLIMMIX | Yes | None | 0.922 | 0.029 | | |
| GLIMMIX+ SIMEX | Yes | SIMEX (500) | 0.938 | 0.085 | 16 | 447 |
| GLIMMIX * Adj1 | Yes | Eq 12 | 0.937 | 0.034 | 64 | 4 |
| GLIMMIX * Adj2 | Yes | Eq 13 | 0.950 | 0.034 | 64 | 4 |
| GLIMMIX * Adj2a | Yes | Eq 14 | 0.926 | 0.034 | 64 | 4 |
| GLIMMIX * Adj3 | Yes | Eq 15 | 0.932 | 0.034 | 64 | 4 |

Notes:

## 4.2 Example Application

We applied the methods discussed in this paper to the evaluation of a federal grants program called the Health Profession Opportunity Grants (HPOG) Program.[26] These grants are given to community colleges, workforce development boards, and other organizations to provide education, training, and support services to Temporary Assistance for Needy Families (TANF) recipients and other low-income adults for occupations in the healthcare field that pay well and are expected to either experience labor shortages or be in high demand. There have been two rounds of grants. We worked with data from the federally-sponsored impact evaluation of the first round of HPOG (referred to as HPOG 1.0).[27] The evaluation included 23 organizations that had been awarded grants, which then developed and implemented 42 programs. Under the impact evaluation, about 13,800 applicants were randomized to HPOG or business as usual. There were follow-up surveys at 15 months, three years, and six years. We used data from the first two of these follow-up surveys. Excluding nonrespondents at either round as well as respondents from three grantees that were co-enrolled in a parallel evaluation with a slightly different set of questionnaires, we were able to use a total sample of about 6,600 participants at 37 programs.

[26] HPOG was authorized by the Affordable Care Act (ACA), Public Law 111-148, 124 Stat. 119, March 23, 2010, sect. 5507(a), “Demonstration Projects to Provide Low-Income Individuals with Opportunities for Education, Training, and Career Advancement to Address Health Professions Workforce Needs,” adding sect. 2008(a) to the Social Security Act, 42 U.S.C. 1397g(a). The Administration for Children and Families within the U.S. Department of Health and Human Services administers the HPOG grants and federal research and evaluation portfolio.

Our general approach was to use variables collected at 15 months to measure potential mediating program features and to use variables collected at three years as outcomes. We examined a fairly large set of mediators and outcomes. Most of these failed to indicate any mediation. Two exceptions involved mediation by program-average FTE months of study in the month of random assignment and the following six months. At the person level, the average value was 2.6 months with a standard deviation of 2.7 months and an ICC of 0.09. The average sample size per program was 178 with a relative variance of $V_n^2$=0.55. Using equation (12), these parameters should result in an estimate by GLIMMIX of the mediation coefficient that requires about 11 percent inflation.

[27] Walton, Harvill and Peck (2019) reported on the mediation of several variables on local impact estimates at 15 months using this same study, but their methodology was more like that of BHR than the methodology discussed in this paper. All of the mediators they studied were based on two surveys of people involved in administering the programs rather than of people experiencing the programs. One of the surveys they used was of grantees and a second of management and staff. They also did not use control arm mean outcomes as moderators, instead they used local data from federal statistical agencies and baseline profiles of students such as the percent without a high diploma. Also, they used search techniques to identify mediators.

The two outcomes where mediation by this variable appears to be active are:

- Subjective perception of career progress at interview
- Earning a degree or a certificate that requires at least a year of study as of 36 months

Table 3 shows the results in a format parallel to Table 2. For both outcomes, the naïve model fails to detect any mediation. However, once the BLUP for the control arm mean outcome is introduced as a covariate, a significant mediation effect on the second outcome emerges. Furthermore, once the estimates are corrected for attenuation with any of the asymptotic adjustments, significant mediation effects appear for both outcomes. The SIMEX adjustments are smaller and have larger standard errors, so if one used SIMEX, one would not detect this mediation. While this mediation seems plausible (programs that enroll their student quickly and keep them in training longer have larger impacts on sense of career progress and longer term credentials at three years), these findings must be viewed with skepticism because of the hunt across outcomes and potential mediators,. The main point of showing these findings here is to demonstrate that the technique has the potential to uncover mediation that would otherwise remain hidden.

Table 3. Illustration of Methods on HPOG 1.0 Sample

| Outcome and Method | CBLUP in Model? | Method for Correction of Attenuation | $\hat{\lambda}$ (Mediating effect of Z) | SE | Jack Reps | CPU Minutes |
|---|---|---|---|---|---|---|
| Mediation of program-average FTE months of study by month 6 on sense of career progress at three years | | | | | | |
| Naïve model | No | None | -.013 | .008 | | |
| GLIMMIX | Yes | None | .042 | .022 | | |
| GLIMMIX+ SIMEX | Yes | SIMEX (500) | .045 | .023 | 16 | 22 |
| GLIMMIX * Adj1 | Yes | Eq 12 | .046* | .020 | 42 | <1 |
| GLIMMIX * Adj2 | Yes | Eq 13 | .049* | .020 | 42 | <1 |
| GLIMMIX * Adj2a | Yes | Eq 14 | .046* | .020 | 42 | <1 |
| GLIMMIX * Adj3 | Yes | Eq 15 | .046* | .020 | 42 | <1 |
| Mediation of program-average FTE months of study by month 6 on earning (by year 3) a degree or a certificate requiring at least one year of study | | | | | | |
| Naïve model | No | None | -.008 | .005 | | |
| GLIMMIX | Yes | None | .041* | .018 | | |
| GLIMMIX+ SIMEX | Yes | SIMEX (500) | .041 | .027 | 16 | 30 |
| GLIMMIX * Adj1 | Yes | Eq 12 | .046* | .018 | 42 | <1 |
| GLIMMIX * Adj2 | Yes | Eq 13 | .049* | .020 | 42 | <1 |
| GLIMMIX * Adj2a | Yes | Eq 14 | .046* | .018 | 42 | <1 |
| GLIMMIX * Adj3 | Yes | Eq 15 | .046* | .018 | 42 | <1 |

* Statistically significant at .05 level.

## 5. Discussion

This paper has laid out a method for using aggregate participant data in an ecometric approach to both control for local contextual factors and measure local potential site-level mediators in support of BHR-style exploration of the sources of cross-site impact variance in multi-site randomized studies. These methods have focused on a single outcome and a single mediator. As suggested by a referee, these methods could be generalized to remove confounding due to cross-site variation in outcomes other than the focal outcome. For example, in a study like HPOG, one could add BLUPS of the control arm means of education outcomes when studying labor force outcomes and vice versa. Regarding parallel mediation, it should also be possible to add the means of treatment-arm assessments of multiple dimensions of the student experience to the reduced-form model in equation (9).

This paper suggests a jackknife approach to variance estimation. The authors also have studied linearization estimators, but this approach is more difficult. We hope to report on them in a future paper. Linearized forms are not needed for variance estimation, but they would be useful to support power projection during the design phase of large-scale multi-site evaluations.

## References

Aguinis, H. and Culpepper, S.A. (2015). "An Expanded Decision-Making Procedure for Examining Cross-level Interaction Effects with Multilevel Modeling." *Organization Research Methods* 18: 155-176.

Ammann, C.M., Genton, M.G., and Li, B. (2010). "Technical Notes: Correcting for Signal Attenuation from Noisy Proxy Data in Climate Reconstructions." *Climate of the Past* 6: 273-279.

Anderson, G.J. (1970). "Effects of Classroom Social Climate on Individual Learning." *American Education Research Journal* 7: 135-152.

Asparouhov, T. and Muthen, B. (2007). "Constructing Covariates in Multilevel Regression." *Mplus Web Notes*: No 11.

Bloom, H. S. (ed.). 2005. *Learning More from Social Experiments: Evolving Analytic Approaches.* New York: Russell Sage Foundation.

Bloom, H. S., Hill, C.J., Riccio, J.A. 2003. "Linking program implementation and effectiveness: Lessons from a pooled sample of welfare-to-work experiments." *Journal of Policy Analysis and Management* 22: 551-75.

Braga, M., Paccagnella, M., Pellizzari (2014). Evaluating students' evaluation of professors. *Economics of Education Review*, 41, 71-88.

Carrell, S.E. and West, J.E. (2010). Does professor quality matter? Evidence from random assignment of students to professors. *Journal of Political Economy*, 118, 409-432.

Cook, J.R., Stefanski, L.A. 1994. "Simulation-extrapolation estimation in parametric measurement error models." *Journal of the American Statistical Association* 89: 1314-1328.

Goldstein, H. 2011. *Multilevel Statistical Models*. Chicester: John Wiley.

Creemers, B. P., Kyriakides, L., Sammons, P. 2010. *Methodological Advances in Educational Effectiveness Research*. Oxon: Routledge.

Davis, T.J. (2015). *Teacher Beliefs and Perceptions about Preschool Bullying.* PhD Dissertation, University of Alabama.

Efron, Bradley; Stein, Charles (May 1981). "The Jackknife Estimate of Variance". *The Annals of Statistics*. **9** (3): 586–596. doi:10.1214/aos/1176345462

Hedges, L.V. and Hedberg, E.C. (2007). "Intraclass correlation values for planning group-randomized trials in education. *Educational Evaluation and Policy Analysis* 29: 60-87.

Hox, J.J. 2010. *Multilevel Analysis*: *Techniques and Applications*. New York: Routledge.

James Bell and Associates (2009). *Evaluation Brief: Measuring Implementation Fidelity*. Arlington, VA: James Bell and Associates. http://www.acf.hhs.gov/sites/default/files/cb/measuring_implementation_fidelity.pdf

Johnson, A.D., Markowitz, A.J., Hill, C.J., and Philipps, D.A. (2016). “Variation in Impacts of Tulsa Pre-K on Cognitive Development in Kindergarten: The Role of Instructional Support.” *Developmental Psychology* 52: 2145-2158.

Li, H., Lahiri, P. 2010. "An adjusted maximum likelihood method for solving small area estimation problems." *Journal of Multivariate Analysis* 101: 882-892.

Lipsky, M. 1980. *Street-Level Bureaucracy: Dilemmas of the Individual in Public Services.* New York, NY: Russel Sage Foundation.

Mackenbach, Joreintje D., Jeroen Lakerveld, Frank J. van Lenthe, Ichiro Kawachi, Martin McKee, Harry Rutter, Ketevan Glonti et al. "Neighbourhood social capital: measurement issues and associations with health outcomes." *obesity reviews* 17 (2016): 96-107.

Malone, L. (2015). *My Existence Didn't Make No Difference to Them: Perceptions of Teacher Expectations Among African-American Students and Their Families.* (Electronic Thesis or Dissertation). Retrieved from https://etd.ohiolink.edu/

Marsh, H.W., Lüdtke, O., Nagengast, B., Trautwein, U., Morin, A.J.S., Abduljabbar, A.A., and Köller, O. (2012). “Classroom Climate and Contextual Effects: Conceptual and Methodological Issues in the Evaluation of Group-Level Effects.” *Education Psychologist* 47: 103-124.

Martin, Gina, Joanna Inchley, Gerry Humphris, and Candace Currie. "Assessing the psychometric and ecometric properties of neighborhood scales using adolescent survey data from urban and rural Scotland." *Population health metrics* 15, no. 1 (2017): 11. Moerbeek, M., Teerenstra, S. 2016. *Power Analysis of Trials with Multilevel Data*. Boca Raton: CRC Press.

Morris, Pamela A., Maia Connors, Allison Friedman-Krauss, Dana Charles McCoy, Christina Weiland, Avi Feller, Lindsay Page, Howard Bloom, and Hirokazu Yoshikawa. "New findings on impact variation from the Head Start Impact Study: Informing the scale-up of early childhood programs." *AERA Open* 4, no. 2 (2018): 2332858418769287.

Peck, Laura R., Alan Werner, Eleanor Harvill, Daniel Litwok, Shawn Moulton, Alyssa Rulf Fountain, and Gretchen Locke. (2018). *Health Profession Opportunity Grants (HPOG 1.0) Impact Study Interim Report: Program Implementation and Short-Term Impacts*, OPRE Report 2018-16a. Washington, DC: Office of Planning, Research, and Evaluation, Administration for Children and Families, U.S. Department of Health and Human Services.

Peck, L. R., Werner, W., Fountain, A. R., Buell, J. L., Bell, S. H., Harvill, E., Nisar, H., Judkins, D., Locke, G. 2014. *Health Profession Opportunity Grants Impact Study Design Report*. OPRE Report #2014-62, Washington, DC: Office of Planning Research and Evaluation, Administration for Children and Families, U.S. Department of Health and Human Services.

Raudenbush, Stephen W., and Robert J. Sampson. 1999. Ecometrics: Toward a Science of Assessing Ecological Settings, with Application to the Systematic Social Observation of Neighborhoods. *Sociological Methodology*. 29(1):1-41. https://journals.sagepub.com/doi/10.1111/0081-1750.00059

Reardon, Sean F. and Stephen W. Raudenbush. 2013. "Under What Assumptions Do Site-by-Treatment Instruments Identify Average Causal Effects?" *Sociological Methods and Research* 42(2): 143-163.

Reardon, Sean F., Fatih Unlu, Pei Zhu, and Howard S. Bloom. 2014. "Bias and Bias Correction in Multisite Instrumental Variables Analysis of Heterogeneous Mediator Effects." *Journal of Educational and Behavioral Statistics* 39(1): 53-86.

Robinson, G. K. 1991. "That BLUP is a Good Thing: The Estimation of Random Effects." *Statistical Science* 6(1): 15-32.

Rothstein, J., & Von Wachter, T. (2017). Social experiments in the labor market. In *Handbook of economic field experiments* (Vol. 2, pp. 555-637). North-Holland.

Schunck, R. (2016). "Cluster Size and Aggregated Level 2 Variables in Multilevel Models: A Cautionary Note." *Methods, Data, Analyses* 10: 97-108.

Stefanski, L.A. and Cook, J.R. (1995). Simulation-extrapolation: The Measurement error jackknife. *Journal of the American Statistical Association* 90: 1247-1256.

Smith, Jeffrey Andrew and Whalley, Alexander and Wilcox, Nathaniel, Are Program Participants Good Evaluators?. IZA Discussion Paper No. 13584, Available at SSRN: https://ssrn.com/abstract=3674305

Tukey, John W. (1958). "Bias and confidence in not quite large samples (abstract)". *The Annals of Mathematical Statistics*. **29** (2): 614. doi:10.1214/aoms/1177706647

Valliant, R., Dorfman, A. H., Royall, R. M. 2000. *Finite Population Sampling and Inference: A Prediction Approach*. New York: John Wiley & Sons.

Walton, Douglas, Eleanor L. Harvill, and Laura R. Peck (2019). *Which Program Characteristics Are Linked to Program Impacts? Lessons from the HPOG 1.0 Evaluation.* OPRE Report 2019- 51, Washington, DC: Office of Planning, Research, and Evaluation, Administration for Children and Families, U.S. Department of Health and Human Services.

Weiss, M. J., Bloom, H. S., & Brock, T. (2014). A conceptual framework for studying the sources of variation in program effects. *Journal of Policy Analysis and Management*, *33*(3), 778-808.

Weiss, Michael J., Howard S. Bloom, Natalya Verbitsky-Savitz, Himani Gupta, Alma E. Vigil, and Daniel N. Cullinan. "How much do the effects of education and training programs vary across sites? Evidence from past multisite randomized trials." *Journal of Research on Educational Effectiveness* 10, no. 4 (2017): 843-876.

Wolter, K.M. (2007). *Introduction to Variance Estimation* (2nd ed.). New York: Springer.

## Appendix: Derivation of Approximate Disattenuation Coefficient

In this appendix, we approximate the attenuation in the estimate of the mediation coefficient when following the suggestion in Section 3.2, we estimate in by fitting the following reduced model at the site level using SAS?GLIMMIX:

$$\hat{\mu}_j = \delta + \hat{u}_j \varphi + L_j \eta + \bar{z}_{.jT} \lambda + q_j + g_j \qquad (9)$$

where and $q_j \sim \mathrm{N}\left(0, \tau_q^2\right)$ and $g_j \sim N\left(0, 4\sigma^2 / n_j\right)$ are iid errrors, independent of each other, and $\sigma^2$ is assumed to be known. As a first step, we rewrite this model in matrix notation.

Let $\mathbf{H} = \begin{bmatrix} 1 & \bar{z}_{.1T} \\ \vdots & \vdots \\ 1 & \bar{z}_{.mT} \end{bmatrix}$, $\mathbf{V} = \begin{bmatrix} \tau_q^2 + 4\sigma^2 / n_1 & & 0 \\ & \ddots & \\ 0 & & \tau_q^2 + 4\sigma^2 / n_m \end{bmatrix}$, $\mathbf{W} = \mathbf{V}^{-1}$, $\mathbf{L} = \begin{bmatrix} \hat{\mu}_1 \\ \vdots \\ \hat{\mu}_m \end{bmatrix}$, $\mathbf{A} = \mathbf{H}'\mathbf{W}\mathbf{H}$, and $\mathbf{\Psi} = \begin{bmatrix} \delta \\ \lambda \end{bmatrix}$.

Since $\tau_q^2$ is unknown, it is not possible to use **W** as a weight matrix in fitting a linear model for **L** in terms of **H**, but asymptotically, the use of SAS/GLIMMIX suggested in the main body of the paper approximates this.

For a general weight matrix **W**, then

$$\hat{\Psi}_{WLS} = \mathbf{A}^{-1}\mathbf{H}'\mathbf{W}\mathbf{L} \text{ and } \mathrm{var}\left(\hat{\Psi}_{WLS}\right) = \mathbf{A}^{-1}\mathbf{H}'\mathbf{W}\mathbf{V}\mathbf{W}\mathbf{H}\mathbf{A}^{-1}$$

If we were able to use $\mathbf{W} = \mathbf{V}^{-1}$ as the weight matrix, this would simplify to

$\hat{\Psi}_{WLS} = \mathbf{A}^{-1}\mathbf{H}'\mathbf{V}^{-1}\mathbf{L}$ and $\operatorname{var}\left(\hat{\Psi}_{WLS}\right) = \mathbf{A}^{-1}\mathbf{H}'\mathbf{V}^{-1}\mathbf{H}\mathbf{A}^{-1} = \mathbf{A}^{-1}$

Working through the linear algebra,

$$\mathbf{A}^{-1} = \frac{1}{\left(\sum w_j\right)\left(\sum w_j \bar{z}_{.jT}^2\right) - \left(\sum w_j \bar{z}_{.jT}\right)^2} \begin{bmatrix} \sum w_j \bar{z}_{.jT}^2 & -\sum w_j \bar{z}_{.jT} \\ -\sum w_j \bar{z}_{.jT} & \sum w_j \end{bmatrix},$$

$$\hat{\lambda} = \frac{\left(\sum w_j\right)\left(\sum w_j \bar{z}_{.jT} \hat{\mu}_j\right) - \left(\sum w_j \bar{z}_{.jT}\right)\left(\sum w_j \hat{\mu}_j\right)}{\left(\sum w_j\right)\left(\sum w_j \bar{z}_{.jT}^2\right) - \left(\sum w_j \bar{z}_{.jT}\right)^2},$$

And

$$\operatorname{var}\hat{\lambda} = \frac{\sum w_j}{\left(\sum w_j\right)\left(\sum w_j \bar{z}_{.jT}^2\right) - \left(\sum w_j \bar{z}_{.jT}\right)^2},$$

where $w_j = \dfrac{1}{\tau_q^2 + 4\sigma^2 / n_j}$.

The expected value of this ratio may be approximated as the ratio of expected values:

$$\mathrm{E}\hat{\lambda} \approx \frac{\mathrm{E}\left[\left(\sum w_j\right)\left(\sum w_j \bar{z}_{.jT}\hat{\mu}_j\right)\right] - \mathrm{E}\left[\left(\sum w_j \bar{z}_{.jT}\right)\left(\sum w_j \hat{\mu}_j\right)\right]}{\mathrm{E}\left[\left(\sum w_j\right)\left(\sum w_j \bar{z}_{.jT}^2\right)\right] - \mathrm{E}\left(\sum w_j \bar{z}_{.jT}\right)^2}$$

$$= \frac{m\mathrm{cov}\left(w_{j*}, w_{j*}\bar{z}_{.j*T}\hat{\mu}_{j*}\right) + m^2\mathrm{E}\left(w_{j*}\right)\mathrm{E}\left(w_{j*}\bar{z}_{.j*T}\hat{\mu}_{j*}\right) - m\mathrm{cov}\left(w_{j*}\bar{z}_{.j*T}, w_{j*}\hat{\mu}_{j*}\right) - m^2\mathrm{E}\left(w_{j*}\bar{z}_{.j*T}\right)\mathrm{E}\left(w_{j*}\hat{\mu}_{j*}\right)}{m\mathrm{cov}\left(w_{j*}, w_{j*}\bar{z}_{.j*T}^2\right) + m^2\mathrm{E}\left(w_{j*}\right)\mathrm{E}\left(w_{j*}\bar{z}_{.j*T}^2\right) - m\mathrm{var}\left(w_{j*}\bar{z}_{.j*T}\right) - m^2\mathrm{E}^2\left(w_{j*}\bar{z}_{.j*T}\right)}$$

$$= \frac{\mathrm{cov}\left(w_{j*}, w_{j*}\bar{z}_{.j*T}\hat{\mu}_{j*}\right)/m + \mathrm{E}\left(w_{j*}\right)\mathrm{E}\left(w_{j*}\bar{z}_{.j*T}\hat{\mu}_{j*}\right) - \mathrm{cov}\left(w_{j*}\bar{z}_{.j*T}, w_{j*}\hat{\mu}_{j*}\right)/m - \mathrm{E}\left(w_{j*}\bar{z}_{.j*T}\right)\mathrm{E}\left(w_{j*}\hat{\mu}_{j*}\right)}{\mathrm{cov}\left(w_{j*}, w_{j*}\bar{z}_{.j*T}^2\right)/m + \mathrm{E}\left(w_{j*}\right)\mathrm{E}\left(w_{j*}\bar{z}_{.j*T}^2\right) - \mathrm{var}\left(w_{j*}\bar{z}_{.j*T}\right)/m - \mathrm{E}^2\left(w_{j*}\bar{z}_{.j*T}\right)}$$

where $j*$ is a randomly selected site.

Assuming that the site sample sizes are non-informative so that

$$\mathrm{E}\left(w_{j*}\bar{z}_{.j*T}\right) = \mathrm{E}\left(w_{j*}\right)\mathrm{E}\left(\bar{z}_{.j*T}\right) = 0,$$

$$\mathrm{cov}\left(w_{j*}, w_{j*}\bar{z}_{.j*T}\hat{\mu}_{j*}\right) = \mathrm{E}\left(\bar{z}_{.j*T}\hat{\mu}_{j*}\right)\mathrm{var}\left(w_{j*}\right),$$

$$\mathrm{E}\left(w_{j*}\bar{z}_{.j*T}\hat{\mu}_{j*}\right) = \mathrm{E}\left(w_{j*}\right)\mathrm{E}\left(\bar{z}_{.j*T}\hat{\mu}_{j*}\right)$$

$$\mathrm{cov}\left(w_{j*}\bar{z}_{.j*T}, w_{j*}\hat{\mu}_{j*}\right) = \mathrm{cov}\left(\bar{z}_{.j*T}, \hat{\mu}_{j*}\right)\left[\mathrm{var}\left(w_{j*}\right) + \mathrm{E}^2\left(w_{j*}\right)\right]$$

We can re-express the approximated expected value of $\hat{\lambda}$ as

$$\mathrm{E}\hat{\lambda} \approx \frac{\operatorname{cov}\left(w_{j*}, w_{j*}\bar{z}_{.j*T}\hat{\mu}_{j*}\right)/m + \mathrm{E}\left(w_{j*}\right)\mathrm{E}\left(w_{j*}\bar{z}_{.j*T}\hat{\mu}_{j*}\right) - \operatorname{cov}\left(w_{j*}\bar{z}_{.j*T}, w_{j*}\hat{\mu}_{j*}\right)/m - \mathrm{E}\left(w_{j*}\bar{z}_{.j*T}\right)\mathrm{E}\left(w_{j*}\hat{\mu}_{j*}\right)}{\operatorname{cov}\left(w_{j*}, w_{j*}\bar{z}^2_{.j*T}\right)/m + \mathrm{E}\left(w_{j*}\right)\mathrm{E}\left(w_{j*}\bar{z}^2_{.j*T}\right) - \operatorname{var}\left(w_{j*}\bar{z}_{.j*T}\right)/m - \mathrm{E}^2\left(w_{j*}\bar{z}_{.j*T}\right)}$$

$$= \frac{\mathrm{E}\left(\bar{z}_{.j*T}\hat{\mu}_{j*}\right)\operatorname{var}\left(w_{j*}\right)/m + \mathrm{E}^2\left(w_{j*}\right)\mathrm{E}\left(\bar{z}_{.j*T}\hat{\mu}_{j*}\right) - \operatorname{cov}\left(\bar{z}_{.j*T}, \hat{\mu}_{j*}\right)\left[\operatorname{var}\left(w_{j*}\right) + \mathrm{E}^2\left(w_{j*}\right)\right]/m}{\operatorname{cov}\left(w_{j*}, w_{j*}\bar{z}^2_{.j*T}\right)/m + \mathrm{E}\left(w_{j*}\right)\mathrm{E}\left(w_{j*}\bar{z}^2_{.j*T}\right) - \operatorname{var}\left(w_{j*}\bar{z}_{.j*T}\right)/m}$$

$$= \frac{\left\{\operatorname{var}\left(w_{j*}\right)\mathrm{E}\left(\bar{z}_{.j*T}\hat{\mu}_{j*}\right) - \left[\operatorname{var}\left(w_{j*}\right) + \mathrm{E}^2\left(w_{j*}\right)\right]\operatorname{cov}\left(\bar{z}_{.j*T}, \hat{\mu}_{j*}\right)\right\}/m + \mathrm{E}^2\left(w_{j*}\right)\mathrm{E}\left(\bar{z}_{.j*T}\hat{\mu}_{j*}\right)}{\left[\operatorname{cov}\left(w_{j*}, w_{j*}\bar{z}^2_{.j*T}\right) - \operatorname{var}\left(w_{j*}\bar{z}_{.j*T}\right)\right]/m + \mathrm{E}\left(w_{j*}\right)\mathrm{E}\left(w_{j*}\bar{z}^2_{.j*T}\right)}$$

We now deal separately with each of the terms in this expression with the trick of conditioning on $j*$ and making use of the facts that under model (1) with $\varphi = \eta = 0$, $\mathrm{E}\left(\hat{\mu}_{j*} \middle| j*\right) = \delta + Z_{j*}\lambda + q_{j*}$, $\mathrm{E}Z_{j*} = 0$. Also, we assume that the local sample sizes are independent of the local implementation quality and effects, so that $\operatorname{cov}\left(w_{j*}, Z_{j*}\right) = \operatorname{cov}\left(w_{j*}, q_{j*}\right) = 0$.

We first deal with terms that do not involve the weights:

$$\begin{aligned}\mathrm{E}\left(\bar{z}_{.j*T}\hat{\mu}_{j*}\right) &= \mathrm{E}\left[\mathrm{E}\left(\bar{z}_{.j*T}\hat{\mu}_{j*} \middle| j*\right)\right] \\ &= \mathrm{E}\left[Z_{j*}\left(\delta + Z_{j*}\lambda + q_{j*}\right)\right] \\ &= \lambda\tau_z^2\end{aligned}$$

$$\operatorname{cov}\left(\bar{z}_{.j^*T}, \hat{\mu}_{j^*}\right) = \operatorname{cov}\left[\mathrm{E}\left(w_{j^*}\left(Z_{j^*} + \varepsilon_{Zj^*}\right) \middle| j^*\right), \mathrm{E}\left(w_{j^*}\left(\delta + Z_{j^*}\lambda + q_{j^*} + \bar{e}_{.j^*T} - \bar{e}_{.j^*C}\right) \middle| j^*\right)\right]$$

$$= \operatorname{cov}\left(Z_{j^*}, \left(\delta + Z_{j^*}\lambda + q_{j^*}\right)\right)$$

$$= \lambda \operatorname{var}\left(Z_{j^*}\right) = \lambda \tau_Z^2$$

$$\operatorname{var}\left(\bar{z}_{.j^*T}\right) = \mathrm{E}\left[\operatorname{var}\left(\bar{z}_{.j^*T} \middle| j^*\right)\right] + \operatorname{var}\left[\mathrm{E}\left(\bar{z}_{.j^*T} \middle| j^*\right)\right]$$

$$= \mathrm{E}\left[2\tau_z^2\left(1-\rho_z\right)/\rho_z/n_{j^*}\right] + \operatorname{var}\left[Z_{j^*}\right]$$

$$= \left[2\tau_z^2\left(1-\rho_z\right)/\rho_z/\tilde{n}\right] + \tau_z^2$$

$$= \tau_z^2\left[2\left(1-\rho_z\right)/\rho_z/\tilde{n} + 1\right]$$

and

$$\mathrm{E}\left(\bar{z}_{.j^*T}^2\right) = \mathrm{E}\left(\bar{z}_{.j^*T}^2\right)$$

$$= \mathrm{E}\left[\mathrm{E}\left(\bar{z}_{.j^*T}^2 \middle| j^*\right)\right]$$

$$= \mathrm{E}\left[\mathrm{Z}_j^2 + \operatorname{var}\left(\bar{z}_{.j^*T} \middle| j^*\right)\right]$$

$$= \left\{\tau_z^2 + \mathrm{E}\left[2\tau_z^2\left(1-\rho_z\right)/\rho_z/n_{j^*}\right]\right\}$$

$$= \tau_z^2\left[1 + 2\left(1-\rho_z\right)/\rho_z/\tilde{n}\right]$$

Where $\rho_Z$ is the intraclass correlation in Z, and $\tilde{n}$ is the harmonic average of the site sample sizes.

We now deal with the terms involving weights and the squared values of the mediators. These are complex because although the weights are independent of the mediators, they are not independent of the local variance of the estimated mediator.

$$
\begin{aligned}
\operatorname{cov}\left(w_{j*}, w_{j*}\bar{z}^2_{.j*T}\right) &= \mathrm{E}\left[\operatorname{cov}\left(w_{j*}, w_{j*}\bar{z}^2_{.j*T} \,|\, j*\right)\right] + \operatorname{cov}\left[\mathrm{E}\left(w_{j*} \,|\, j*\right), \mathrm{E}\left(w_{j*}\bar{z}^2_{.j*T} \,|\, j*\right)\right] \\
&= 0 + \operatorname{cov}\left(w_{j*}, w_{j*}\left[\operatorname{var}\left(\bar{z}_{.j*T} \,|\, j*\right) + \mathrm{E}^2\left(\bar{z}_{.j*T} \,|\, j*\right)\right]\right) \\
&= \operatorname{cov}\left(w_{j*}, w_{j*}\left[2\tau_z^2\left(1-\rho_z\right)/\rho_z/n_{j*} + Z^2_{j*}\right]\right) \\
&= \operatorname{cov}\left(w_{j*}, w_{j*}2\tau_z^2\left(1-\rho_z\right)/\rho_z/n_{j*}\right) + \operatorname{cov}\left(w_{j*}, w_{j*}Z^2_{j*}\right) \\
&= \left[2\tau_z^2\left(1-\rho_z\right)/\rho_z\right]\operatorname{cov}\left(w_{j*}, w_{j*}/n_{j*}\right) + \operatorname{var}\left(w_{j*}\right)\tau_Z^2
\end{aligned}
$$

and

$$
\begin{aligned}
\mathrm{E}\left(w_{j*}\bar{z}^2_{.j*T}\right) &= \mathrm{E}\left[\mathrm{E}\left(w_{j*}\bar{z}^2_{.j*T} \,|\, j*\right)\right] \\
&= \mathrm{E}\left[w_{j*}\mathrm{E}\left(\bar{z}^2_{.j*T} \,|\, j*\right)\right] \\
&= \mathrm{E}\left[w_{j*}\mathrm{Z}_j^2 + w_{j*}\operatorname{var}\left(\bar{z}_{.j*T} \,|\, j*\right)\right] \\
&= \mathrm{E}\left(w_{j*}\right)\tau_z^2 + \mathrm{E}\left[w_{j*}2\tau_z^2\left(1-\rho_z\right)/\rho_z/n_{j*}\right] \\
&= \mathrm{E}\left(w_{j*}\right)\tau_z^2 + 2\tau_z^2\mathrm{E}\left(w_{j*}/n_{j*}\right)\left(1-\rho_z\right)/\rho_z
\end{aligned}
$$

and

$$
\begin{aligned}
\operatorname{var}\left(w_{j*}\bar{z}_{.j*T}\right) &= \mathrm{E}\left[\operatorname{var}\left(w_{j*}\bar{z}_{.j*T} \mid j*\right)\right] + \operatorname{var}\left[\mathrm{E}\left(w_{j*}\bar{z}_{.j*T} \mid j*\right)\right] \\
&= \mathrm{E}\left[2\tau_z^2 w_{j*}^2 \left(1-\rho_z\right)/\rho_z / n_{j*}\right] + \operatorname{var}\left[w_{j*} Z_{j*}\right] \\
&= 2\tau_z^2\left(1-\rho_z\right)/\rho_z \mathrm{E}\left(w_{j*}^2 / n_{j*}\right) + \tau_z^2 \mathrm{E}\left(w_{j*}^2\right) \\
&= 2\tau_z^2\left(1-\rho_z\right)/\rho_z \mathrm{E}\left(w_{j*}^2 / n_{j*}\right) + \tau_z^2\left[\operatorname{var}\left(w_{j*}^2\right) + \mathrm{E}^2\left(w_{j*}\right)\right]
\end{aligned}
$$

So,

$$
\begin{aligned}
\operatorname{cov}\left(w_{j*}, w_{j*}\bar{z}_{.j*T}^2\right) - \operatorname{var}\left(w_{j*}\bar{z}_{.j*T}\right) &= \left[2\tau_z^2\left(1-\rho_z\right)/\rho_z\right]\operatorname{cov}\left(w_{j*}, w_{j*}/n_{j*}\right) + \operatorname{var}\left(w_{j*}\right)\tau_Z^2 + \\
&\quad -\left[2\tau_z^2\left(1-\rho_z\right)/\rho_z\right]\mathrm{E}\left(w_{j*}^2/n_{j*}\right) - \tau_z^2\left[\operatorname{var}\left(w_{j*}^2\right) + \mathrm{E}^2\left(w_{j*}\right)\right] \\
&= \left[2\tau_z^2\left(1-\rho_z\right)/\rho_z\right]\left[\operatorname{cov}\left(w_{j*}, w_{j*}/n_{j*}\right) - \mathrm{E}\left(w_{j*}^2/n_{j*}\right)\right] - \tau_z^2\mathrm{E}^2\left(w_{j*}\right) \\
&= -\left[2\tau_z^2\left(1-\rho_z\right)/\rho_z\right]\left[\mathrm{E}\left(w_{j*}\right)\mathrm{E}\left(w_{j*}/n_{j*}\right)\right] + \tau_z^2\mathrm{E}^2\left(w_{j*}\right)
\end{aligned}
$$

Combining these results, we get the new approximation that

$$\mathrm{E}\hat{\lambda} \approx \frac{\left\{\operatorname{var}\left(w_{j*}\right)\mathrm{E}\left(\bar{z}_{.j*T}\hat{\mu}_{j*}\right)-\left[\operatorname{var}\left(w_{j*}\right)+\mathrm{E}^2\left(w_{j*}\right)\right]\operatorname{cov}\left(\bar{z}_{.j*T},\hat{\mu}_{j*}\right)\right\}/m+\mathrm{E}^2\left(w_{j*}\right)\mathrm{E}\left(\bar{z}_{.j*T}\hat{\mu}_{j*}\right)}{\left[\operatorname{cov}\left(w_{j*},w_{j*}\bar{z}^2_{.j*T}\right)-\operatorname{var}\left(w_{j*}\bar{z}_{.j*T}\right)\right]/m+\mathrm{E}\left(w_{j*}\right)\mathrm{E}\left(w_{j*}\bar{z}^2_{.j*T}\right)}$$

$$=\frac{\left\{\operatorname{var}\left(w_{j*}\right)\lambda\tau_z^2-\left[\operatorname{var}\left(w_{j*}\right)+\mathrm{E}^2\left(w_{j*}\right)\right]\lambda\tau_Z^2\right\}/m+\mathrm{E}^2\left(w_{j*}\right)\lambda\tau_z^2}{\left\{\left[2\tau_z^2\left(1-\rho_z\right)/\rho_z\right]\left[\mathrm{E}\left(w_{j*}\right)\mathrm{E}\left(w_{j*}/n_{j*}\right)\right]-\tau_z^2\mathrm{E}^2\left(w_{j*}\right)\right\}/m+\mathrm{E}\left(w_{j*}\right)\left[\mathrm{E}\left(w_{j*}\right)\tau_z^2+2\tau_z^2\mathrm{E}\left(w_{j*}/n_{j*}\right)\left(1-\rho_z\right)/\rho_z\right]}$$

$$=\frac{\lambda\tau_z^2\mathrm{E}^2\left(w_{j*}\right)\left(m-1\right)/m}{\left\{\left[2\tau_z^2\left(1-\rho_z\right)/\rho_z\right]\left[\mathrm{E}\left(w_{j*}\right)\mathrm{E}\left(w_{j*}/n_{j*}\right)\right]+\tau_z^2\mathrm{E}^2\left(w_{j*}\right)\right\}\left(m-1\right)/m}$$

$$=\frac{\lambda\mathrm{E}\left(w_{j*}\right)}{\mathrm{E}\left(w_{j*}\right)+\left[2\left(1-\rho_z\right)/\rho_z\right]\left[\mathrm{E}\left(w_{j*}/n_{j*}\right)\right]}$$

$$=\frac{\lambda}{1+\left[2\left(1-\rho_z\right)/\rho_z\right]\left[\mathrm{E}\left(w_{j*}/n_{j*}\right)\right]/\mathrm{E}\left(w_{j*}\right)}$$

We now turn to the terms involving only the weights and sample sizes. First, a third-order Taylor series approximation to $E\left(w_{j*}\right)$ is

$$\mathrm{E}\left(w_{j*}\right)\approx\mathrm{E}\left[\frac{\bar{n}}{\bar{n}\tau_q^2+4\sigma^2}+\frac{4\sigma^2\left(n_{j*}-\bar{n}\right)}{\left(\bar{n}\tau_q^2+4\sigma^2\right)^2}-\frac{4\tau_q^2\sigma^2\left(n_{j*}-\bar{n}\right)^2}{\left(\bar{n}\tau_q^2+4\sigma^2\right)^3}+\frac{4\tau_q^4\sigma^2\left(n_{j*}-\bar{n}\right)^3}{\left(\bar{n}\tau_q^2+4\sigma^2\right)^4}\right]$$

$$=\frac{\bar{n}}{\bar{n}\tau_q^2+4\sigma^2}+0-\frac{4\tau_q^2\sigma^2\operatorname{var}n_{j*}}{\left(\bar{n}\tau_q^2+4\sigma^2\right)^3}+\frac{4\tau_q^4\sigma^2\operatorname{skew}\left(n_{j*}\right)\left(\operatorname{var}n_{j*}\right)^{\frac{3}{2}}}{\left(\bar{n}\tau_q^2+4\sigma^2\right)^4}$$

$$=\frac{1}{\bar{n}\tau_q^2+4\sigma^2}\left[\bar{n}-\frac{4\tau_q^2\sigma^2\operatorname{var}n_{j*}}{\left(\bar{n}\tau_q^2+4\sigma^2\right)^2}+\frac{4\tau_q^4\sigma^2\operatorname{skew}\left(n_{j*}\right)\left(\operatorname{var}n_{j*}\right)^{\frac{3}{2}}}{\left(\bar{n}\tau_q^2+4\sigma^2\right)^3}\right]$$

Similarly, a third-order Taylor series approximation to $\mathrm{E}\left(w_{j*}/n_{j*}\right)$ is

$$E_*\left(w_{j*}/n_{j*}\right) \approx E_*\left[\frac{1}{\bar{n}\tau_q^2+4\sigma^2}-\frac{\tau_q^2\left(n_{j*}-\bar{n}\right)}{\left(\bar{n}\tau_q^2+4\sigma^2\right)^2}+\frac{\tau_q^4\left(n_{j*}-\bar{n}\right)^2}{\left(\bar{n}\tau_q^2+4\sigma^2\right)^3}-\frac{\tau_q^6\left(n_{j*}-\bar{n}\right)^3}{\left(\bar{n}\tau_q^2+4\sigma^2\right)^4}\right]$$

$$=\frac{1}{\bar{n}\tau_q^2+4\sigma^2}\left[1+\frac{\tau_q^4\operatorname{var} n_{j*}}{\left(\bar{n}\tau_q^2+4\sigma^2\right)^2}-\frac{\tau_q^6\operatorname{skew}\left(n_{j*}\right)\left(\operatorname{var} n_{j*}\right)^{\frac{3}{2}}}{\left(\bar{n}\tau_q^2+4\sigma^2\right)^3}\right]$$

Putting these together, the expected attenuation factor can be approximated as

$$\frac{\mathrm{E}\hat{\lambda}}{\lambda} \approx \frac{1}{1+\left[2\left(1-\rho_z\right)/\rho_z\right]\left[\mathrm{E}\left(w_{j*}/n_{j*}\right)\right]/\mathrm{E}\left(w_{j*}\right)}$$

$$=\frac{1}{1+\left[2\left(1-\rho_z\right)/\rho_z\right]\left[1+\frac{\tau_q^4\operatorname{var}_* n_{j*}}{\left(\bar{n}\tau_q^2+4\sigma^2\right)^2}-\frac{\tau_q^6\operatorname{skew}\left(n_{j*}\right)\left(\operatorname{var} n_{j*}\right)^{\frac{3}{2}}}{\left(\bar{n}\tau_q^2+4\sigma^2\right)^3}\right]\Bigg/\left[\bar{n}-\frac{4\tau_q^2\sigma^2\operatorname{var}_* n_{j*}}{\left(\bar{n}\tau_q^2+4\sigma^2\right)^2}+\frac{4\tau_q^4\sigma^2\operatorname{skew}\left(n_{j*}\right)\left(\operatorname{var} n_{j*}\right)^{\frac{3}{2}}}{\left(\bar{n}\tau_q^2+4\sigma^2\right)^3}\right]}$$

$$=\frac{1}{1+\left[2\left(1-\rho_z\right)/\rho_z\right]\left[\left(\tau_q^2+4\sigma^2/\bar{n}\right)^3+\tau_q^4\left(\tau_q^2+4\sigma^2/\bar{n}\right)V_n^2-\tau_q^6\operatorname{skew}\left(n_{j*}\right)V_n^3\right]\Big/\left[\bar{n}\left(\tau_q^2+4\sigma^2/\bar{n}\right)^3-4\tau_q^2\sigma^2\left(\tau_q^2+4\sigma^2/\bar{n}\right)V_n^2+4\tau_q^4\sigma^2\operatorname{skew}\left(n_{j*}\right)V_n^3\right]}$$

where $V_n^2=\operatorname{var} n_{j*}/\bar{n}^2$ is the relative variance in the site-level sample sizes.

If the sample sizes are distributed as a gamma distribution (as seems generally plausible), then $\operatorname{skew}\left(n_{j*}\right)=2\sqrt{V_n^2}$ , in which case, the expected attenuation factor is:

$$\frac{\mathrm{E}\hat{\lambda}}{\lambda} \approx$$

$$\frac{1}{1+\left[2\left(1-\rho_z\right)/\rho_z\right]\left[\left(\tau_q^2+4\sigma^2/\bar{n}\right)^3+\tau_q^4\left(\tau_q^2+4\sigma^2/\bar{n}\right)V_n^2-\tau_q^6\mathrm{skew}\left(n_{j*}\right)V_n^3\right]\Big/\left[\bar{n}\left(\tau_q^2+4\sigma^2/\bar{n}\right)^3-4\tau_q^2\sigma^2\left(\tau_q^2+4\sigma^2/\bar{n}\right)V_n^2+4\tau_q^4\sigma^2\mathrm{skew}\left(n_{j*}\right)V_n^3\right]}$$

$$=\frac{1}{1+\left[2\left(1-\rho_z\right)/\rho_z\right]\left[\left(\tau_q^2+4\sigma^2/\bar{n}\right)^3+\tau_q^4\left(\tau_q^2+4\sigma^2/\bar{n}\right)V_n^2-\tau_q^6 2\sqrt{V_n^2}V_n^3\right]\Big/\left[\bar{n}\left(\tau_q^2+4\sigma^2/\bar{n}\right)^3-4\tau_q^2\sigma^2\left(\tau_q^2+4\sigma^2/\bar{n}\right)V_n^2+4\tau_q^4\sigma^2 2\sqrt{V_n^2}V_n^3\right]}$$

$$=\frac{1}{1+\left[2\left(1-\rho_z\right)/\rho_z\right]\left[\left(\tau_q^2+4\sigma^2/\bar{n}\right)^3+\tau_q^4\left(\tau_q^2+4\sigma^2/\bar{n}\right)V_n^2-2\tau_q^6V_n^4\right]\Big/\left[\bar{n}\left(\tau_q^2+4\sigma^2/\bar{n}\right)^3-4\tau_q^2\sigma^2\left(\tau_q^2+4\sigma^2/\bar{n}\right)V_n^2+8\tau_q^4\sigma^2V_n^4\right]}$$

Inverting this leads to the adjustment in equation (15) of the main text.

Further simplification is possible under various extreme conditions. First, it is clear that as either $\rho_z \to 1$ or $\bar{n} \to \infty$, the attenuation disappears. We explore limiting values of this attenuation factor under three other conditions: $\tau_q^2 \to 0$, $\tau_q^2 \to \infty$, and $V_n^2 \to 0$.

Second, leading to the adjustment in equation (12) of the main text,

$$\lim_{\tau_q^2 \to 0} \frac{\mathrm{E}\hat{\lambda}}{\lambda} \approx \lim_{\tau_q^2 \to 0} \frac{1}{1+\left[2\left(1-\rho_z\right)/\rho_z\right]\left[\left(\tau_q^2+4\sigma^2/\overline{n}\right)^3+\tau_q^4 V_n^2\left(\tau_q^2+4\sigma^2/\overline{n}\right)-2\tau_q^6 V_n^4\right]\Big/\left[\overline{n}\left(\tau_q^2+4\sigma^2/\overline{n}\right)^3-4\tau_q^2\sigma^2 V_n^2\left(\tau_q^2+4\sigma^2/\overline{n}\right)+8\tau_q^4\sigma^2 V_n^4\right]}$$

$$= \lim_{\tau_q^2 \to 0} \frac{1}{1+\left[2\left(1-\rho_z\right)/\rho_z\right]\left[\left(4\sigma^2/\overline{n}\right)^3\right]\Big/\left[\overline{n}\left(4\sigma^2/\overline{n}\right)^3\right]}$$

$$= \frac{1}{1+\left[2\left(1-\rho_z\right)/\rho_z\right]/\overline{n}}$$

$$= \frac{\rho_z\overline{n}}{2\left(1-\rho_z\right)+\rho_z\overline{n}}$$

Third, leading to the adjustment in equation (14) of the main text,

$$\lim_{\tau_q^2 \to \infty} \frac{\mathrm{E}\hat{\lambda}}{\lambda} \approx \lim_{\tau_q^2 \to \infty} \frac{1}{1+\left[2\left(1-\rho_z\right)/\rho_z\right]\left[\left(\tau_q^2+4\sigma^2/\overline{n}\right)^3+\tau_q^4 V_n^2\left(\tau_q^2+4\sigma^2/\overline{n}\right)-2\tau_q^6 V_n^4\right]\Big/\left[\overline{n}\left(\tau_q^2+4\sigma^2/\overline{n}\right)^3-4\tau_q^2\sigma^2 V_n^2\left(\tau_q^2+4\sigma^2/\overline{n}\right)+8\tau_q^4\sigma^2 V_n^4\right]}$$

$$= \lim_{\tau_q^2 \to \infty} \frac{1}{1+\left[2\left(1-\rho_z\right)/\rho_z\right]\left[1+\tau_q^4 V_n^2/\left(\tau_q^2+4\sigma^2/\overline{n}\right)^2-2\tau_q^6 V_n^4/\left(\tau_q^2+4\sigma^2/\overline{n}\right)^3\right]\Big/\left[\overline{n}-4\tau_q^2\sigma^2 V_n^2/\left(\tau_q^2+4\sigma^2/\overline{n}\right)^2+8\tau_q^4\sigma^2 V_n^4/\left(\tau_q^2+4\sigma^2/\overline{n}\right)^3\right]}$$

$$= \lim_{\tau_q^2 \to \infty} \frac{1}{1+\left[2\left(1-\rho_z\right)/\rho_z\right]\left[1+V_n^2/\left(1+4\sigma^2/\tau_q^2/\overline{n}\right)^2-2V_n^4/\left(1+4\sigma^2/\tau_q^2/\overline{n}\right)^3\right]\Big/\left[\overline{n}-4\sigma^2 V_n^2/\left(\tau_q+4\sigma^2/\tau_q/\overline{n}\right)^2+8\sigma^2 V_n^4/\left(\tau_q^{2/3}+4\sigma^2/\tau_q^{2/3}/\overline{n}\right)^3\right]}$$

$$= \frac{1}{1+\left[2\left(1-\rho_z\right)/\rho_z\right]\left(1+V_n^2-2V_n^4\right)/\overline{n}}$$

If $V_n^2 < 1$, the adjustment can be simplified as in equation (13) of the main text as follows:

$$\lim_{\tau_q^2 \to \infty, V_n^2 < 1} \frac{\mathrm{E}\hat{\lambda}}{\lambda} \approx \frac{1}{1 + \left[2\left(1-\rho_z\right)/\rho_z\right]\left(1+V_n^2\right)/\overline{n}}$$
$$= \frac{\rho_z \overline{n}}{\rho_z \overline{n} + \left[2\left(1-\rho_z\right)\right]\left(1+V_n^2\right)}$$

Fourth, also leading to the adjustment in equation (12) of the main text:

$$\lim_{V_n^2 \to 0} \frac{\mathrm{E}\hat{\lambda}}{\lambda} \approx \lim_{V_n^2 \to 0} \frac{1}{1 + \left[2\left(1-\rho_z\right)/\rho_z\right]\left[\left(\tau_q^2 + 4\sigma^2/\overline{n}\right)^3 + \tau_q^4 V_n^2\left(\tau_q^2 + 4\sigma^2/\overline{n}\right) - 2\tau_q^6 V_n^4\right] \Big/ \left[\overline{n}\left(\tau_q^2 + 4\sigma^2/\overline{n}\right)^3 - 4\tau_q^2\sigma^2 V_n^2\left(\tau_q^2 + 4\sigma^2/\overline{n}\right) + 8\tau_q^4\sigma^2 V_n^4\right]}$$
$$= \frac{\rho_z \overline{n}}{2\left(1-\rho_z\right) + \rho_z \overline{n}}$$

Lastly, we note that it is not possible with this approximation to estimate the limiting value for infinite relative variance in site-level sample sizes,[28] but it seems safe to assume that the second-order approximation will work better than the first-order approximation.

[28] Just taking the limit of this approximation results in "attenuation" factors greater than one, which seems clearly impossible. Even though the skew of a gamma variable can be ignored in this approximation, the kurtosis cannot since it is of the same magnitude as the relative variance.

## Materials for On-line Supplement

```
* Mediation by average student feedback.sas;
* Dave Judkins;
* November 29, 2018;

* This program implements an ecometric approach to BHR-style analyses with a
single mediator;

%macro GenerateSeeds(MasterSeed,NumSeeds);
    data Seeds;
      array Seeds (&NumSeeds);
      masterseed = &masterseed;
      do i = 1 to &NumSeeds;
        call ranuni(MasterSeed, Seeds[i]);
        Seeds[i] = int(100000000*Seeds[i]);
        end;
      drop i;
      dummy = 1;
      output;
    run;
%mend GenerateSeeds;

%macro AddSuffix(InList,Suffix);
  rename
  %let j=1;
  %let eof=0;
  %do %while (&EOF=0);
    %let OrigName=%scan(&InList,&j,' ');
    %if %length(&OrigName)>0 %then %do;
        &OrigName=&OrigName._&Suffix
        %let j=%eval(&j+1);
      %end;
      %else %do;
        %let eof=1;
      %end;
  %end;
  ;
%mend AddSuffix;

%macro GenSiteVarList(InList,Suffix,OutListName);
  %global &OutListName;
  %Let _TempList=;
  %let j=1;
  %let eof=0;
  %do %while (&EOF=0);
    %let OrigName=%scan(&InList,&j,' ');
    %if %length(&OrigName)>0 %then %do;
        %let _TempList=&_TempList &OrigName._&Suffix;
        %let j=%eval(&j+1);
      %end;
      %else %do;
        %let eof=1;
      %end;
  %end;
```

```
  %let &OutListName=&_TempList;
%mend GenSiteVarList;

%macro GroupCenter(InList,Suffix1,Suffix2);
  %let j=1;
  %let eof=0;
  %do %while (&EOF=0);
    %let OrigName=%scan(&InList,&j,' ');
    %if %length(&OrigName)>0 %then %do;
        &OrigName._&Suffix1=&OrigName-&OrigName._&Suffix2;
        %let j=%eval(&j+1);
      %end;
      %else %do;
        %let eof=1;
      %end;
  %end;
%mend GroupCenter;

%macro CBLUP(InDSN,OutDSN,Treatment,Level2Unit,Outcome,CovList,TitleLevel);
  * This program calculates BLUPS of control arm means to use as a level-two
moderator;
  * It assumes that treatment status is coded as 1/2 or 1/0 with 1=treated;
  Data _Controls;
    set &InDSN;
      if &Treatment~=1;
  run;
  Title&TitleLevel. "BLUPs of control subject mean outcomes";
  ods graphics off;
  proc mixed data=_Controls;
    class &Level2Unit;
      model &Outcome=&CovList;
      random intercept/subject=&Level2Unit solution;
      ods output SolutionR=&OutDSN;
  run;
  ods graphics on;
  data &OutDSN;
    set &OutDSN;
      rename estimate=CBLUP;
      keep &Level2Unit estimate;
  run;
%mend CBLUP;

%Macro
LocalEffects(InDSN,OutDSN,Treatment,Level2Unit,Outcome,Level1CovList,TitleLev
el);
  Title&TitleLevel. "Calculation of local effects";
  ods graphics off;

  proc surveyreg data=&InDSN;
    class &Level2Unit;
      model &Outcome=&Level1CovList &Level2Unit &Treatment*&Level2Unit/
solution;
      /*Note omission of main effect for T. That simplified calculation of
variances on local effects*/
      ods output ParameterEstimates=_Lparms;
  run;
  ods graphics on;
```

```
  proc summary data=&InDSN noprint;
    var &outcome;
      class &Level2Unit;
      ways 1;
      output out=_level2ss n=ss;
  run;
  data _level2ss;
    set _level2ss;
      Evar=4/ss;
      keep &Level2Unit Evar;
  run;
  Data _Lparms;
    set _LParms;
    &Level2Unit=1*scan(parameter,2,' ');
    if index(parameter,"&Treatment*")>0;
    LocalEffectVar=stderr**2;
    rename estimate=LocalEffect;
    keep &Level2Unit estimate LocalEffectVar;
  run;
  Data &OutDSN;
    merge _Lparms _level2ss;
      by &Level2unit;
      LocalEffectVar=(LocalEffectVar+Evar)/2; /*stabilize residual variances
on local effects using the expected variance without covariates*/
    keep &Level2Unit LocalEffect LocalEffectVar;
  run;
%mend LocalEffects;

%macro Mediator(InDSN,OutDSN,Treatment,Level2Unit,Mediator);
  proc summary data=&InDSN noprint;
    class &Level2Unit;
      var &Mediator;
      ways 1;
      output out=&OutDSN mean=Level2Mediator stderr=Level2MedStdErr;
      where &Treatment=1;
  run;
  Data &OutDSN;
    set &OutDSN;
      Level2MedVar=Level2MedStdErr**2;
      dummy=1;
      keep &Level2Unit Level2Mediator Level2MedVar dummy;
  run;
%mend Mediator;

%macro
DilutedMediation(LocalEffects,Level2Moderators,CBLUPS,Level2MediatorFile,OutD
SN,Level2Unit,LocalEffect,LocalEffectVar,
      Level2ModeratorList,Level2Mediator,TitleLevel,MainOrRep);
  * This macro uses the trick of Mukhopadhyay and McDowell to estimate the
mediation model;
  %if "&MainOrRep"="M" %then %let PCONV=1e-6;
  %else %if "&MainOrRep"="R" %then %let PCONV=1e-3;
  Data _g2;
    set &LocalEffects;
      row=_n_;
      col=_n_;
      value=LocalEffectVar;
```

```
      parm=1;
      keep row col value parm;
  run;
  Data _InModel;
    merge &LocalEffects &Level2Moderators &CBLUPS &Level2MediatorFile;
      by &Level2Unit;
  run;
  Title&TitleLevel. "Estimate diluted mediation";
  %if %length(&CBLUPS)>0 %then %do;
    %let Unconfounder=CBLUP;
  %end;
  %else %do;
    %let Unconfounder=;
  %end;
  proc glimmix data=_InModel PCONV=&PCONV method=rspl;
    class &Level2Unit;
    model LocalEffect=&Level2ModeratorList &Unconfounder
&Level2Mediator/solution;
      random &Level2Unit;
    random _residual_ /type =lin(1) ldata=_g2;
      ods output ParameterEstimates=FixedParms
ConvergenceStatus=GLIMMIXConverge CovParm=TauQSq;
  run;

  proc sql noprint;
    select status into: Success from GLIMMIXConverge;
  quit;

  %if &Success=0 %then %do;
    %if %SysFunc(Fileexist(TauQSq)) %then %do;
      Data TauQSq;
          set TauQsq;
          if CovParm="&Level2Unit";
          rename estimate=TauQsqHat;
          dummy=1;
        run;
      %end;

      %else %do;
        Data TauQsq;
          TauQsqhat=0;
            dummy=1;
        run;
      %end;

    Data FixedParms;
      set FixedParms;
        if effect="&Level2Mediator";
        rename estimate=DilutedEffect
           stdErr=DilutedEffect_StdErr;
        dummy=1;
    run;

      data &OutDSN;
        merge FixedParms TauQSq;
        by dummy;
      run;
```

```
  %end;
  %else %do;
    Data &OutDSN;
        DilutedEffect=.;
      DilutedEffect_StdErr=.;
        TauQsqHat=.;
        dummy=1;
    run;
  %end;
%mend DilutedMediation;

%macro
SIMEXMediation(LocalEffects,Level2Moderators,CBLUPS,Level2MediatorFile,OutDSN
,Level2Unit,LocalEffect,LocalEffectVar,Level2ModeratorList,
  Level2Mediator,TitleLevel,Increment,NumPoints,seed,MainOrRep);
  *This macro adds random noise to the level-two mediator, and then
extrapolates to zero measurement error;
      %if "&MainOrRep"="M" %then %do; /*(Already turned off if MainOrRep=R)*/
        ods select none;
        options nonotes;
        ods results off;
      %end;
  %GenerateSeeds(&seed,&NumPoints);
  %Let NextTitle=%eval(&TitleLevel+1);
  Data _CumSIMEX;
    _Lambda=-1;
    _LambdaSq=1;
      DilutedEffect=.;
    output;
  run;
  %let j=1;
  %do i= 1 %to &NumPoints;
    Data _NoisyMediator;
        merge seeds &Level2MediatorFile;
        by dummy;
        _Lambda=&j*&Increment;
        _LambdaSq=_Lambda*_Lambda;
      _Multiplier=sqrt(Level2MedVar*_Lambda);
        call rannor(seeds&j,x);
        _NoisedMediator=&Level2Mediator+x*_Multiplier;
        keep &Level2Unit _Lambda &Level2Mediator _NoisedMediator;
      run;
      proc summary data=_NoisyMediator noprint;
        var _NoisedMediator;
      output out=_VarGrowth var=VarNoisedMediator;
      run;
      %if "&MainOrRep"="M" %then %do;
        %put "Running Simex iter &j";
      %end;

%DilutedMediation(&LocalEffects,&Level2Moderators,&CBLUPS,_NoisyMediator,_Sim
exIter,&Level2Unit,&LocalEffect,&LocalEffectVar,
        &Level2ModeratorList,_NoisedMediator,&NextTitle,R);
      Data _SimexIter;
        merge _SimexIter _VarGrowth;
        SimexIter=&j;
        _Lambda=&j*&Increment;
```

```
        _LambdaSq=_Lambda*_Lambda;
      keep _Lambda _LambdaSq SimexIter DilutedEffect VarNoisedMediator;
      run;
      Data _CumSimex;
        set _CumSimex _SimexIter;
      run;
      %let j=%Eval(&j+1);
  %end;
      %if "&MainOrRep"="M" %then %do;
        ods select all;
        options notes;
        ods results on;
      %end;
  Title&TitleLevel. "SIMEX Extrapolation";
  proc reg data=_CumSimex;
    model DilutedEffect=_Lambda _LambdaSq;
      output out=_SimexExtrap pred=SimexMediatorEffect;
  run;
  %if "&MainOrRep"="M" %then %do;
  Title&TitleLevel. "Check on noising process";
  proc reg data=_CumSimex;
    model VarNoisedMediator=_Lambda;
  run;
  %end;
  data &OutDSN;
    set _SimexExtrap;
      if _Lambda=-1;
      keep SimexMediatorEffect;
  run;
%mend SIMEXMediation;

%macro
JackSIMEX(LocalEffects,Level2Moderators,CBLUPS,Level2MediatorFile,OutDSN,Leve
l2Unit,LocalEffect,LocalEffectVar,Level2ModeratorList,
  Level2Mediator,TitleLevel,Increment,NumPoints,NumReps,seed);
  ods select none;
  ods graphics off;
  options nonotes;
  ods results off;
  proc sql noprint;
    select &Level2Unit into: ProgramList separated by ' ' from &LocalEffects;
      select count(*) into: NumPrograms from &LocalEffects;
  quit;
  %if &NumPrograms<=&NumReps %then %do;
    %let NumReps=&NumPrograms;
  %end;
  %else %do;
    proc surveyselect data=&LocalEffects out=_SampReps method=srs
SampSize=&NumReps seed=&seed;
      run;
      proc sql noprint;
        select &Level2Unit into: ProgramList separated by ' ' from _SampReps;
      quit;
  %end;
  Data _CumJack; delete; run;
  %Let JackTitle=%eval(&TitleLevel+1);
  %let EOF=0;
```

```
  %let jj=1;
  %do %while (&EOF=0);
    %Let Drop=%scan(&ProgramList,&jj,' ');
      %if %length(&Drop)>0 %then %do;
      %put "Running SIMEX jackknife iter &jj";
        Data _d1;
          set &LocalEffects;
            if &Level2Unit~="&Drop";
        run;
        %if %length(&Level2Moderators)>0 %then %do;
          Data _d2;
            set &Level2Moderators;
              if &Level2Unit~="&Drop";
          run;
        %end;
        %else %do;
          Data _d2;
          set _d1;
          keep &Level2Unit;
        run;
        %end;
        Data _d3;
          set &CBLUPS;
            if &Level2Unit~="&Drop";
        run;
        Data _d4;
          set &Level2MediatorFile;
            if &Level2Unit~="&Drop";
        run;
        %let JackSeed=%eval(&Seed+&jj);

%SIMEXMediation(_d1,_d2,_d3,_d4,_JackIter,&Level2Unit,&LocalEffect,&LocalEffe
ctVar,&Level2ModeratorList,
          &Level2Mediator,&JackTitle,&Increment,&NumPoints,&Jackseed,R);
        data _JackIter;
          set _JackIter;
            JackIter=&jj;
        run;
        data _CumJack;
          set _CumJack _JackIter;
      run;
      %let jj=%eval(&jj+1);
      %end;
      %else %do;
        %let EOF=1;
      %end;
  %end;
      proc summary data=_CumJack noprint vardef=df;
        var SimexMediatorEffect;
        output out=&OutDSN mean=JackMean var=JackVar;
      run;
      Data &OutDSN;
        set &OutDSN;
        JackStdErr=(&NumPrograms-1)*sqrt(JackVar/&NumPrograms);
        keep JackMean JackStdErr;
      run;
  ods select all;
```

```
  ods graphics on;
  options notes;
  ods results on;
%mend JackSimex;
%macro Inflator
(InDSN,OutDSN,Treatment,Level2Unit,Mediator,MinICCZ,MaxICCZ,EffectFile,Effect
Name,MainOrRep);
  proc summary data=&InDSN noprint;
    var &mediator;
      class &Level2Unit;
      ways 1;
      where &Treatment=1;
      output out=_Level2n n=nj;
  run;

  data _Level2n;
    set _Level2n;
      Inv_nj=1/nj;
      keep &Level2Unit nj Inv_nj;
  run;

  proc summary data=_Level2n noprint;
    var nj Inv_nj;
    output out=_SampleSizes mean=nbar Inv_nbar var=nvar junk;
  run;

  data _SampleSizes;
    set _SampleSizes;
    nHarmonic=1/Inv_nbar;
    dummy=1;
    rename _freq_=NumSites;
      nRelVar=nvar/nbar**2;
    keep dummy nbar nHarmonic nRelVar _freq_;
  run;

  Title&TitleLevel. "Components of variance of mediator";
  proc glimmix data=&InDSN method=rspl;
    class &Level2Unit;
    where &Treatment=1;
    model &Mediator=;
      random site;
    ods output covparms=_ZTComp;
  run; quit;

  Data _ICCZ;
    set _ZTComp;
    dummy=1;
    retain ZBSigma2;
    if upcase(CovParm)=upcase("&Level2Unit") then ZBSigma2=Estimate;
    else do;
      ZWSigma2=Estimate;
        ZSigma2=ZBSigma2+ZWSigma2;
        if ZWSigma2<0 then ICCZ=&MaxICCZ;
      else ICCZ=max(&MinICCZ,ZBSigma2/ZSigma2);
        drop CovParm Estimate StdErr;
        output;
    end;
```

```
  run;

  Data &OutDSN;
    merge _SampleSizes _ICCZ &EffectFile;
    by dummy;
    Inflator_ols=1+2*(1-ICCZ)/nHarmonic/ICCZ;
    Inflator_wls=1+2*(1-ICCZ)/nbar/ICCZ; /*weighted by sample size*/
      Inflator2=1+2*(1-ICCZ)*(1+nRelvar)/nbar/ICCZ;
      ct=(TauQsqhat+4/nbar)**2;
      Inflator3=1++2*(1-ICCZ)/ICCZ*(1+TauQsqhat**2*nRelvar/ct)/(nbar-
4*TauQsqhat*nRelvar/ct);
      AdjustedEffect1=&EffectName*Inflator_wls;
      AdjustedEffect2=&EffectName*Inflator2;
      AdjustedEffect3=&EffectName*Inflator3;
      drop effect;
  run;

  %if "MainOrRep"="M" %then %do;
    Title&TitleLevel. "Adjusted mediation effect";
      Proc print data=&OutDSN;
      run;
  %end;
%mend Inflator;

%macro
JackInflate(InDSN,OutDSN,Treatment,Level2Unit,Mediator,MinICCZ,MaxICCZ,LocalE
ffects,Level2Moderators,CBLUPS,Level2MediatorFile,
      Level2ModeratorList,Level2Mediator,LocalEffect,LocalEffectVar,EffectFil
e,EffectName,TitleLevel,NumReps,Seed);
  ods select none;
  ods graphics off;
  options nonotes;
  ods results off;
  proc sql noprint;
    select &Level2Unit into: ProgramList separated by ' ' from &LocalEffects;
      select count(*) into: NumPrograms from &LocalEffects;
  quit;
  %if &NumPrograms<=&NumReps %then %do;
    %let NumReps=&NumPrograms;
  %end;
  %else %do;
    proc surveyselect data=&LocalEffects out=_SampReps method=srs
SampSize=&NumReps seed=&seed;
      run;
      proc sql noprint;
        select &Level2Unit into: ProgramList separated by ' ' from _SampReps;
      quit;
  %end;
  Data _CumJackInflate; delete; run;
  %let EOF=0;
  %let jj=1;
  %do %while (&EOF=0);
    %let Drop=%scan(&ProgramList,&jj,' ');
      %if %length(&Drop)>0 %then %do;
        %put "Running Inflation jackknife iter &jj";
        Data _d1;
          set &LocalEffects;
```

```
            if &Level2Unit~="&Drop";
        run;
        %if %length(&Level2Moderators)>0 %then %do;
        Data _d2;
            set &Level2Moderators;
              if &Level2Unit~="&Drop";
          run;
        %end;
        %else %do;
          Data _d2;
          set _d1;
          keep &Level2Unit;
        run;
        %end;
        Data _d3;
          set &CBLUPS;
            if &Level2Unit~="&Drop";
        run;
        Data _d4;
          set &Level2MediatorFile;
            if &Level2Unit~="&Drop";
        run;
      Data _d5;
          set &InDSN;
            if &Level2Unit~="&Drop";
        run;


%DilutedMediation(_d1,_d2,_d3,_d4,_JackDilute,&Level2Unit,&LocalEffect,&Local
EffectVar,
                  &Level2ModeratorList,&Level2Mediator,&TitleLevel,R);
      %Inflator
(_d5,_JackInflate,&Treatment,&Level2Unit,&Mediator,&MinICCZ,&MaxICCZ,_JackDil
ute,DilutedEffect,R);
        data _JackInflate;
          set _JackInflate;
            JackIter=&jj;
        run;
        data _CumJackInflate;
          set _CumJackInflate _JackInflate;
      run;
      %let jj=%eval(&jj+1);
      %end;
      %else %do;
        %let EOF=1;
      %end;
  %end;
  proc summary data=_CumJackInflate noprint vardef=df;
      var AdjustedEffect1 AdjustedEffect2 AdjustedEffect3;
      output out=&OutDSN mean=JackMean1 JackMean2 JackMean3 var=JackVar1
JackVar2 JackVar3;
  run;
  Data &OutDSN;
      set &OutDSN;
      JackStdErr1=(&NumPrograms-1)*sqrt(JackVar1/&NumPrograms);
      JackStdErr2=(&NumPrograms-1)*sqrt(JackVar2/&NumPrograms);
      JackStdErr3=(&NumPrograms-1)*sqrt(JackVar3/&NumPrograms);
```

```
      keep JackMean1 JackMean2 JackMean3 JackStdErr1 JackStdErr2 JackStdErr3;
  run;
  ods select all;
  ods graphics on;
  options notes;
  ods results on;
%mend JackInflate;

%macro
EnvironmentalMediation(InDSN,OutDSN,Treatment,Level2Unit,Outcome,Level1CovLis
t,TitleLevel,Mediator,Level2Moderators,
      Level2ModeratorList,SimexIncrement,NumSimexPoints,NumSIMEXJackReps,NumI
nfJackReps,seed);
  * Group center the level-one covariates;
  %GenSiteVarList(&Level1CovList,Resid,XResidList);
  %put "XresidList=&XResidList";
  %GenSiteVarList(&Level1CovList,Mean,XMeanList);
  %put "XMeanList=&XMeanList";
  %let CBLUPcovList=&Level1CovList &Level2ModeratorList &XMeanList;
  proc summary data=&InDSN;
    var &Level1CovList;
      class &Level2Unit;
    ways 1;
      output out=_XMeans mean=;
  run;
  Data _XMeans;
    set _XMeans;
    %AddSuffix(&Level1CovList,Mean);
  run;
  %if %length(&Level2Moderators)>0 %then %do;
    proc sort data=&Level2Moderators out=_Level2Moderators;
      by &Level2Unit;
    run;
    Data _Level2Moderators;
      merge _Level2Moderators _XMeans (keep=&Level2Unit &XMeanList);
        by &Level2Unit;
    run;
  %end;
  %else %do;
    Data _Level2Moderators;
      set _XMeans (keep=&Level2Unit &XMeanList);
        by &Level2Unit;
    run;
  %end;
  data _Centered;
    merge &INDSN _Level2Moderators;
      by &Level2Unit;
      %GroupCenter(&Level1CovList,Resid,Mean);
  run;

  Title&TitleLevel. "Environmental Mediation of Experiment-Based Local
Effects";
  %let TitleLevel=%eval(&TitleLevel+1);

%CBLUP(_Centered,_CBLUPS,&Treatment,&Level2Unit,&Outcome,&CBLUPcovList,&Title
Level);
```

```
%LocalEffects(_Centered,_LocalEffects,&Treatment,&Level2Unit,&Outcome,&XResid
List,&TitleLevel);
  %Mediator(&InDSN,_Level2MediatorFile,&Treatment,&Level2Unit,&Mediator);
  Title&TitleLevel. "Without control arm means as covariates";

%DilutedMediation(_LocalEffects,_Level2Moderators,,_Level2MediatorFile,_Dilut
edEffects,&Level2Unit,LocalEffect,LocalEffectVar,
      &Level2ModeratorList,Level2Mediator,&TitleLevel,M);
  Title&TitleLevel. "With control arm means as covariates";

%DilutedMediation(_LocalEffects,_Level2Moderators,_CBLUPS,_Level2MediatorFile
,_DilutedEffects,&Level2Unit,LocalEffect,LocalEffectVar,
      &Level2ModeratorList,Level2Mediator,&TitleLevel,M);
  %let TimerStart=%sysfunc(Datetime());

%SIMEXMediation(_LocalEffects,_Level2Moderators,_CBLUPS,_Level2MediatorFile,_
SimexOut,&Level2Unit,LocalEffect,LocalEffectVar,&Level2ModeratorList,
    Level2Mediator,&TitleLevel,&SimexIncrement,&NumSimexPoints,&seed,M);
  data _null_;
    dur=datetime()-&TimerStart;
      put 30*'-' / "Duration of Simex Adjustment (&NumSimexPoints points):"
dur time13.2 / 30*'-';
  run;
  %let TimerStart=%sysfunc(Datetime());

%JackSIMEX(_LocalEffects,&Level2Moderators,_CBLUPS,_Level2MediatorFile,_JackO
ut,&Level2Unit,LocalEffect,LocalEffectVar,&Level2ModeratorList,

Level2Mediator,&TitleLevel,&SimexIncrement,&NumSimexPoints,&NumSIMEXJackReps,
&seed);
  data _null_;
    dur=datetime()-&TimerStart;
      put 30*'-' / "Duration of Jackknife variance estimate
(&NumSIMEXJackReps Reps) for SIMEX-adjustment:" dur time13.2 / 30*'-';
  run;
  %let TimerStart=%sysfunc(Datetime());
  %Inflator
(&InDSN,_InflatorOut,&Treatment,&Level2Unit,&Mediator,0.005,.995,_DilutedEffe
cts,DilutedEffect,M);
  data _null_;
    dur=datetime()-&TimerStart;
      put 30*'-' / 'Duration of approximate disattenuation:' dur time13.2 /
30*'-';
  run;
  %let TimerStart=%sysfunc(Datetime());

%JackInflate(&InDSN,_JackInfOut,&Treatment,&Level2Unit,&Mediator,0.005,.7,_Lo
calEffects,&Level2Moderators,_CBLUPS,_Level2MediatorFile,
      &Level2ModeratorList,Level2Mediator,LocalEffect,LocalEffectVar,_Diluted
Effects,DilutedEffect,&TitleLevel,&NumInfJackReps,&Seed);
  data _null_;
    dur=datetime()-&TimerStart;
      put 30*'-' / "Duration of Jackknife variance estimate (&NumInfJackReps
reps) for approximate disattenuation:" dur time13.2 / 30*'-';
  run;
  %let NTitleLevel=%eval(&TitleLevel+1);
```

```
  Title&TitleLevel. "Diluted estimated effect of &Mediator on local effects";
  proc print data=_DilutedEffects;
  run;
  Title&TitleLevel. "SIMEX-corrected estimated effect of &Mediator on local 
effects";
  Title&NTitleLevel. "SIMEX increment= &SimexIncrement Number of SIMEX 
points=&NumSimexPoints";
  proc print data=_SimexOut;
  run;
  Title&TitleLevel. "Jackknife SIMEX-corrected estimated effect of &Mediator 
on local effects";
  proc print data=_JackOut;
  run;
  Title&TitleLevel. "Simple inflated estimated effect of &Mediator on local 
effects";
  proc print data=_InflatorOut;
  run;
  Title&TitleLevel. "Jackknifed simple inflated estimated effect of &Mediator 
on local effects";
  proc print data=_JackInfOut;
  run;

  data &OutDSN;
    length Method $ 32;
    set _DilutedEffects (in=in1 rename=(DilutedEffect=Effect 
DilutedEffect_StdErr=StdErr) drop=effect DF tvalue Probt)
            _SimexOut(in=in2 rename=(SimexMediatorEffect=Effect))
            _JackOut (in=in3 rename=(JackMean=Effect JackStdErr=StdErr))
            _InflatorOut(in=in41 rename=(AdjustedEffect1=Effect))
            _InflatorOut(in=in42 rename=(AdjustedEffect2=Effect))
            _InflatorOut(in=in43 rename=(AdjustedEffect3=Effect))
            _JackInfOut (in=in5 rename=(JackMean1=Effect JackStdErr1=StdErr))
            _JackInfOut (in=in6 rename=(JackMean2=Effect JackStdErr2=StdErr))
            _JackInfOut (in=in7 rename=(JackMean3=Effect JackStdErr3=StdErr))
;
      if in1 then Method="Diluted";
      else if in2 then Method="SIMEX";
      else if in3 then Method="Jackknifed SIMEX";
      else if in41 then Method="Simple inflator 1";
      else if in42 then Method="Simple inflator 2";
      else if in43 then Method="Simple inflator 3";
      else if in5 then Method="Jackknifed inflator1";
      else if in6 then Method="Jackknifed inflator2";
      else if in7 then Method="Jackknifed inflator3";
      keep method effect stderr;
  run;
  Title&TitleLevel. "All estimated effects of &Mediator on local effects";
  proc print data=&OutDSN;
  run;
%mend EnvironmentalMediation;
```